%% file: main.tex
\documentclass[acmsmall,screen]{acmart}

\setcopyright{none}
\renewcommand\footnotetextcopyrightpermission[1]{}
\acmJournal{TOIS}

\usepackage{bm}
\def\method{StageCF}

\usepackage{multirow}
\usepackage{enumitem}
\usepackage{graphicx}
\usepackage{subcaption}
\usepackage{stfloats}

\usepackage{algorithm}
\usepackage{algorithmic}
\usepackage[algo2e,ruled,linesnumbered]{algorithm2e}

\newcommand{\RR}[1]{#1}
\newcommand{\R}[1]{#1}

\AtBeginDocument{%
  
    }

\begin{document}

\title{Interests Burn-down Diffusion Process for Personalized Collaborative Filtering}

\author{Yifang Qin}
\orcid{0000-0002-7520-8039}
\email{qinyifang@pku.edu.cn}
\affiliation{
    \institution{State Key Laboratory for Multimedia Information Processing, School of Computer Science,  PKU-Anker LLM Lab, Peking University, Beijing}
    \country{China}
    \postcode{100871}
}

\author{Zhaobin Li}
\email{2100012903@stu.pku.edu.cn}
\orcid{0009-0003-3012-2658}
\affiliation{
    \institution{School of EECS, Peking University, Beijing}
    \country{China}
    \postcode{100871}
}

\author{Arisa Watanabe}
\email{2000094818@stu.pku.edu.cn}
\orcid{0009-0003-5157-0038}
\affiliation{
    \institution{School of EECS, Peking University, Beijing}
    \country{China}
    \postcode{100871}
}

\author{Wei Ju}
\email{juwei@scu.edu.cn}
\orcid{0000-0001-9657-951X}
\authornote{Corresponding authors.}
\affiliation{
    \institution{College of Computer Science, Sichuan University, Chengdu}
    \country{China}
    \postcode{610065}
}

\author{Zhiping Xiao}
\authornotemark[1]
\email{patxiao@uw.edu}
\orcid{0000-0002-8583-4789}
\affiliation{
    \institution{Paul G. Allen School of Computer Science and Engineering, University of Washington, Seattle, WA}
    \country{USA}
    \postcode{98195}
}


\author{Ming Zhang}
\authornotemark[1]
\email{mzhang\_cs@pku.edu.cn}
\orcid{0000-0002-9809-3430}
\affiliation{
    \institution{State Key Laboratory for Multimedia Information Processing, School of Computer Science,  PKU-Anker LLM Lab, Peking University, Beijing}
    \country{China}
    \postcode{100871}
}
\renewcommand{\shortauthors}{Qin, et al.}

\input{Content/0-abstract}




\received{22 February 2025}
\received[revised]{28 May 2025}
\received[revised]{17 December 2025}
\received[accepted]{6 May 2026}

\maketitle

\input{Content/1-introduction}

\input{Content/2-relatedwork}
\input{Content/3-preliminary}
\input{Content/4-methodology}
\input{Content/5-experiment}

\input{Content/6-conclusion}

\begin{acks}
The authors are grateful to the anonymous reviewers for critically reading the manuscript and for giving important suggestions to improve their paper. 

This paper is partially supported by grants from the National Key Research and Development Program of China with Grant No. 2023YFC3341203, the National Natural Science Foundation of China (NSFC Grant Number 62276002), and Beijing Natural Science Foundation (QY-23048).
\end{acks}

\bibliographystyle{ACM-Reference-Format}
\bibliography{main}


\end{document}

%% file: Content/0-abstract.tex
\begin{abstract}
  Generative methods have gained widespread attention in Collaborative Filtering (CF) tasks for their ability to produce high-quality personalized samples aligned with users' interests. 
  Among them, diffusion generative models have raised increasing attention in recommendation field.
  Despite that the pioneering efforts have applied the conventional diffusion process to model diffusive user interests, the incongruity between the Gaussian noise and the subtle nature of user's personalized interaction behavior has led to sub-optimal results. To this end, we introduce a specifically-tailored diffusion scheme for interaction systems, namely \textit{the interests burn-down process}. The interests burn-down process delineates the decay of user interests towards candidate items, complemented by its reverse burn-up process that yields personalized recommendation for users. The inherent burn-down nature of this process adeptly models the diffusive user interests, aligning seamlessly with the requirements of CF tasks. We present a novel recommendation method \method{} to illustrate the superiority of this newly proposed diffusion process. Experimental results have demonstrated the effectiveness of \method{} against existing generative and diffusion-based baseline methods. Furthermore, comprehensive studies validate the functionality of interests burn-down process, shedding light on its capacity to generate personalized interactions.
\end{abstract}

%% file: Content/1-introduction.tex
\section{Introduction}

As one of the essential components in modern recommender systems, Collaborative Filtering (CF) task stands out as a popular research topic for addressing the information-overload challenge in web applications. The primary objective of CF is to recommend personalized items based on the interaction history of the users. Due to the intricate nature of user interests, lots of researchers have turned to generative methods \cite{wu2016collaborative,wang2017irgan,liang2018variational}, expecting to leverage the expressiveness of generative models and generate suitable recommendation candidates.

Within a series of generative model paradigms, generating samples via diffusion process have received increasing attention for its promising performance across various downstream tasks. Initially conceived for image generation \cite{li2016lattice,ho2020denoising,song2020score}, diffusion generative models have demonstrated success in diverse fields such as as natural language generation \cite{li2022diffusion} and graph construction \cite{xu2021geodiff,huang2022graphgdp}, where the informative diffused samples help generative models accurately depict the subtle prior distribution of data samples. Training a diffusion-based model involves applying a forward diffusion process to observed data samples, with diffused samples drawn from this process to optimize the denoising capability of a score-estimation network. During sample generation, the score-estimation network is utilized to determine the posterior transition rate in the backward diffusion process, which iteratively generating data samples. The chosen diffusion scheme is expected to have a stationary forward distribution, along with an analytical backward posterior for model training and inference. For instance, the widely adopted Gaussian-Markov diffusion \cite{li2016lattice} assumes the forward and backward diffusion are computed through the incremental addition of Gaussian noises. Meanwhile, there are other diffusion schemes tailored for a variety of downstream scenarios, including Dirichlet diffusion \cite{avdeyev2023dirichlet}, transition diffusion \cite{vignac2022digress}, and Blackout diffusion \cite{santos2023blackout}.

There have been pioneering efforts to incorporate the Gaussian diffusion process into recommendation tasks. Notable examples include diffusion CF models \cite{walker2022recommendation,wang2023diffusion} that directly generates interaction vectors for users and \R{sequential recommendation models \cite{du2023sequential,qin2023diffusion,liu2025flow} that employ latent diffusion process to generate personalized and concrete target representations}. For diffusion-based CF methods, prior studies identified a misalignment between the popular Gaussian diffusion-based scheme and processing interaction systems. CODIGEM \cite{walker2022recommendation} relies solely on the output of the initial backward diffusion step,
while DiffRec \cite{wang2023diffusion} recognizes that direct application of Gaussian diffusion leads to collapsed outputs, prompting a reformulation of the training procedure. As a result, the score estimation model in DiffRec degrades into a denoising autoencoder. \R{These issues suggest that modeling interactions in continuous Euclidean space may obscure the diffusion dynamics needed for effective prediction.}
\R{To address this, recent approaches propose alternative diffusion paths tailored to the structure of user–item interactions. HDRM \cite{yuan2025hyperbolic} introduces a hyperbolic diffusion process to capture anisotropic and hierarchical item relationships, and FlowCF \cite{liu2025flowcf} replaces the Gaussian prior with a Bernoulli-based formulation that better reflects discrete user feedback.}


\begin{figure}[t]
\centering
\begin{subfigure}{0.32\linewidth}
    \includegraphics[width=\linewidth]{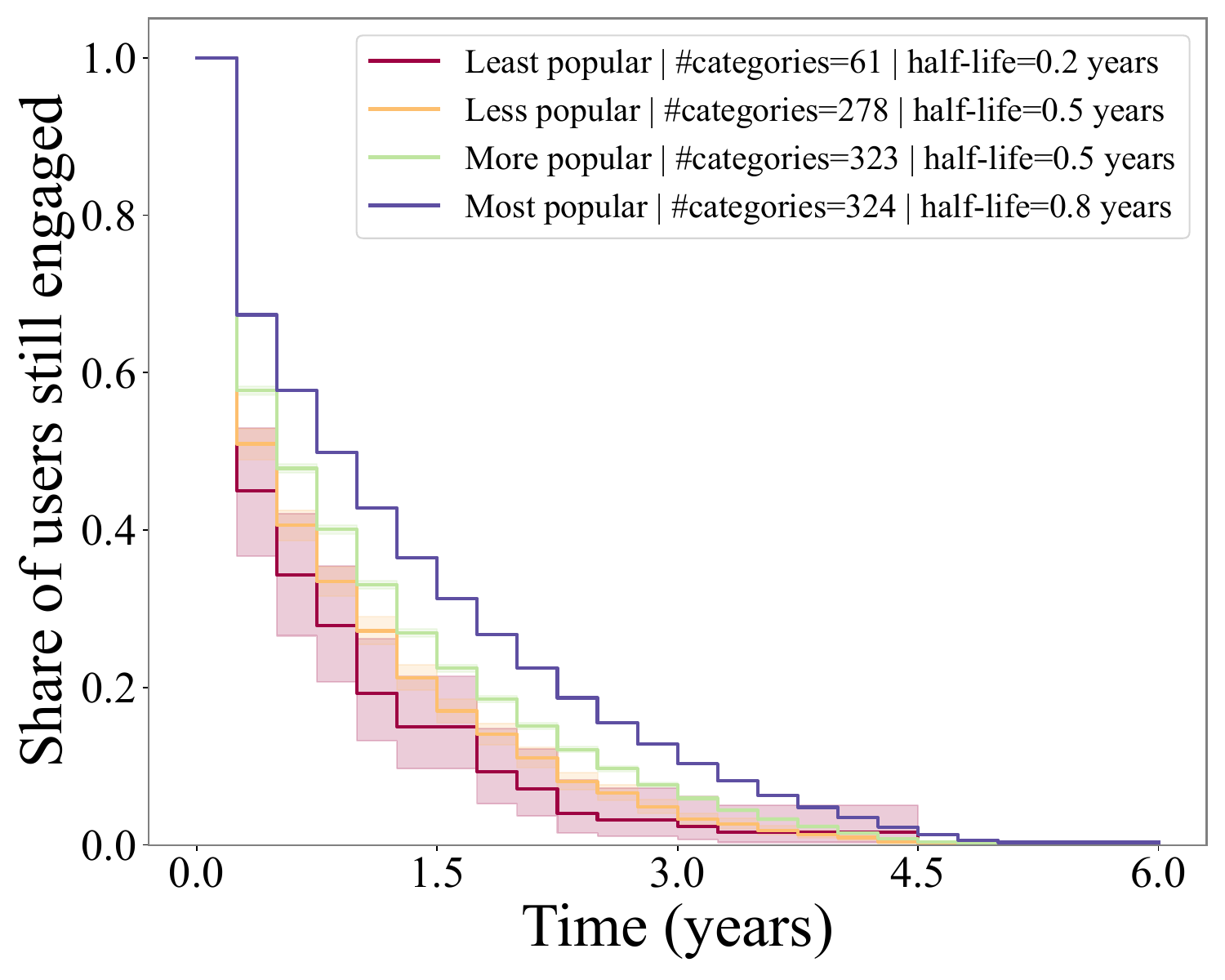}
    \caption{\R{KM curve w.r.t. item popularity.}}
    \label{fig:empirical:a}
\end{subfigure}
\begin{subfigure}{0.32\linewidth}
    \includegraphics[width=\linewidth]{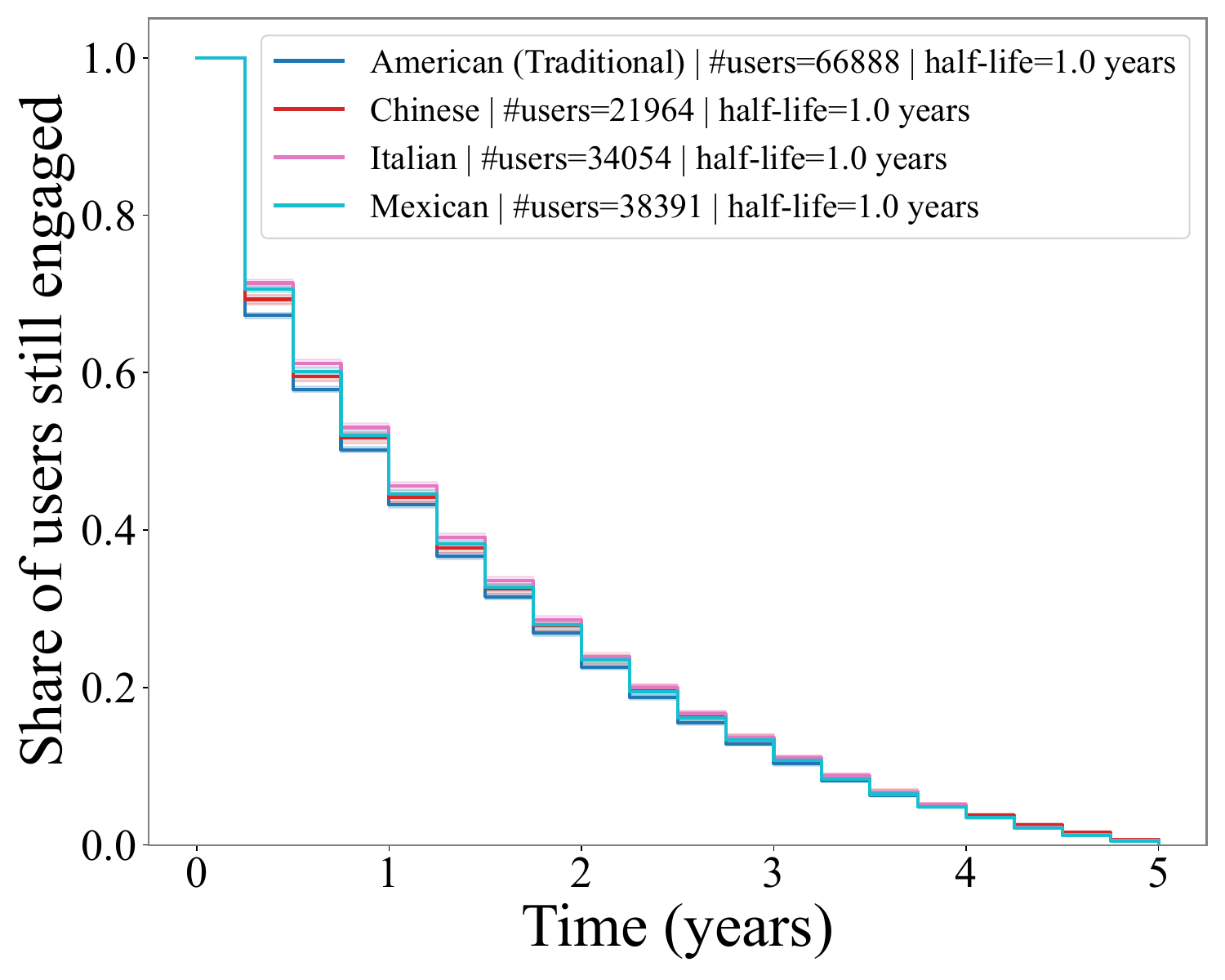}
    \caption{\R{KM curve w.r.t. item category.}}
    \label{fig:empirical:b}
\end{subfigure}
\begin{subfigure}{0.32\linewidth}
    \includegraphics[width=\linewidth]{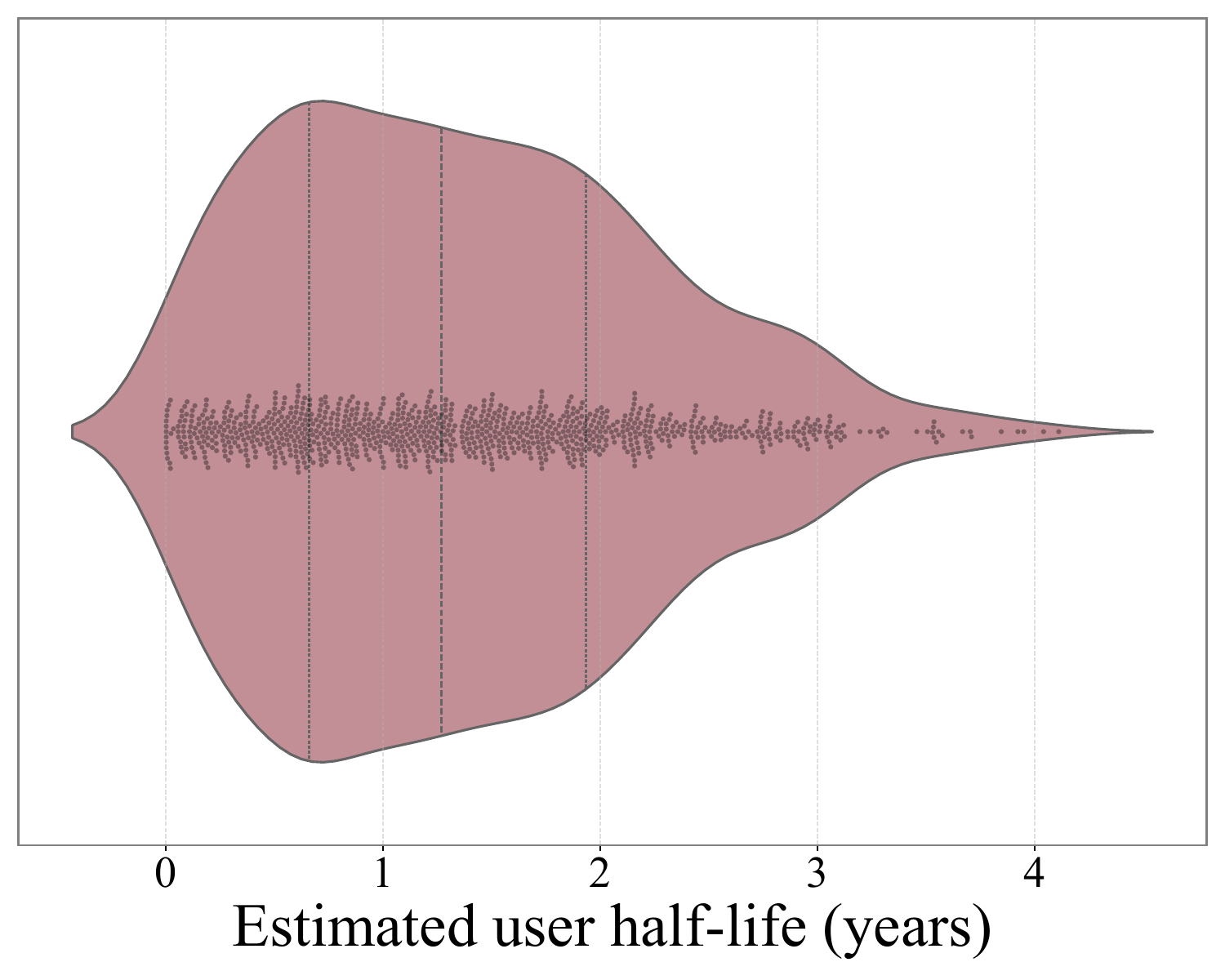}
    \caption{\R{Half-life of user interest.}}
    \label{fig:empirical:c}
\end{subfigure}

\caption{\R{Empirical statistics on the \textbf{Yelp2018} dataset.}} 
\label{fig:empirical}
\end{figure}

\R{Although these advances introduce more suitable diffusion mechanisms, it remains unclear whether these processes preserve key behavioral patterns and fully exploit diffusion models’ ability to infer signals from noisy distributions. To resolve this problem, we conduct a series of empirical studies to identify how user–item interactions evolve over time and to determine a principled way to model this evolution. Using the \textbf{Yelp2018} point-of-interest dataset, we examine user behavioral dynamics across restaurants grouped by popularity and category. For each group, we compute the proportion of returning users and plot Kaplan–Meier survival curves to characterize interest decay. As shown in Figure \ref{fig:empirical:a}, user interest decays more slowly for popular items, whereas Figure \ref{fig:empirical:b} shows that item category has little effect on decay patterns. In addition, the distribution of users’ interest half-lives in Figure \ref{fig:empirical:c} reveals the long-tail variation of users in long-term loyalty. Together, these findings point to an inherent decay trend in user interest, which is shaped strongly by collaborative factors such as item popularity and also influenced by personalized differences across users. 
}

\RR{Drawn from our empirical observations and prior findings, we suggest that an effective diffusion scheme for CF may benefit from preserving certain collaborative patterns in interaction data. Real-world interactions indicate that user interest in items tends to fade over time, suggesting that the forward diffusion process can incorporate a gradual decay mechanism. 
Such a monotonic decay process may offer two potential advantages. First, its intermediate states can reflect varying levels of residual user interest across items, aligning with the heterogeneous decay rates observed in practice. Second, by converging to a stable terminal distribution, the process may provide a more stable training signal for the model.}
\RR{Importantly, this behavior does not only reflect real-world dynamics, but also changes the learning difficulty of different items. By increasing the persistence of low-frequency items during the corruption process, the model is more likely to capture and recover them, which may contribute to improved recommendation performance.}
\RR{In summary, these observations suggest the potential benefits of designing a diffusion scheme that reflects interest evolution and aligns with the characteristics of recommender systems.}


\begin{figure}
\centering
\includegraphics[width=0.85\linewidth]{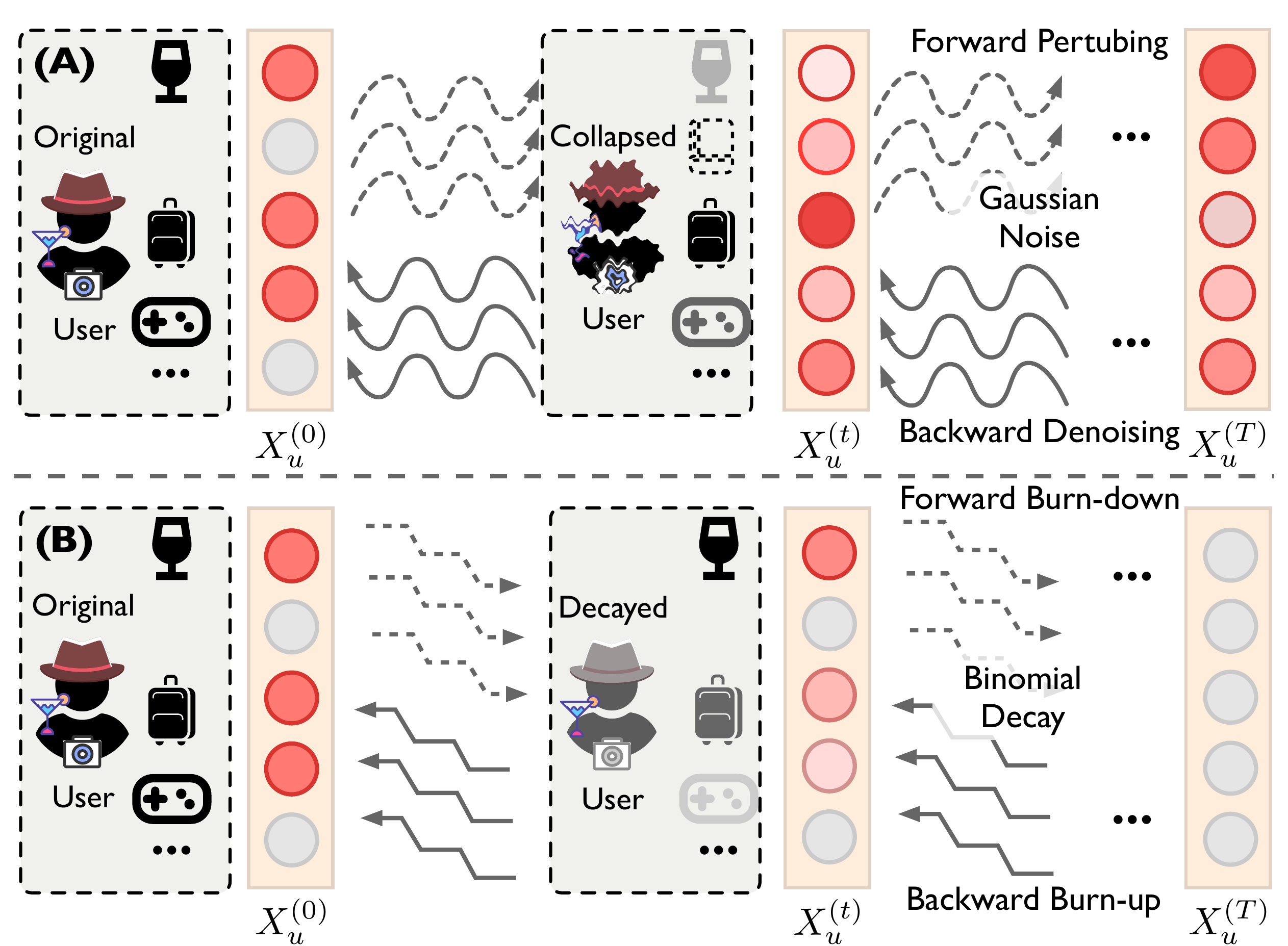}
\caption{An intuitive comparison between (A) a standard Gaussian diffusion process and (B) an interests burn-down process. While Gaussian diffusion eventually leads to model collapse of the user portraits, the interests burn-down process progressively fades the user profile while preserving personalized characteristics.} 
\label{fig:compare}
\end{figure}

\RR{To address the aforementioned challenges, we draw inspiration from} blackout diffusion~\cite{santos2023blackout}, which models discrete pixel data of images through a pure-death process. Building upon this idea, this study introduces the interests burn-down process for personalized recommendations, along with its corresponding model, \method{}. The interests burn-down process captures the stage-wise decay of user interests towards candidate items via a binomial decay process of interest vectors. Meanwhile, the reverse burn-up process is modeled as a incremental sampling process of interests vectors to formulate the final recommendation outcomes. \R{Motivated by our empirical findings that items and users exhibit different decay rates depending on their collaborative attributes, the burn-down process is steered} with a graph convolution factor. Incorporating the collaborative signals enables the burn-up process to generate personalized and behavior-aware recommendations.


\RR{Comprehensive experiments on three widely-used datasets show that the proposed interests burn-down process and \method{} achieve competitive performance.} In summary, our main contributions include:
\begin{itemize}[leftmargin=*]
    \item \textbf{Investigating} the combination of diffusion generative process and CF models, and proposing a specifically designed diffusion scheme, namely the interests burn-down process, tailored for depicting interacting systems. An illustrative comparison of the proposed interests burn-down process and Gaussian diffusion process is presented in Figure \ref{fig:compare}.
    \item \textbf{Building} upon the proposed interests burn-down process to develop a personalized generative CF framework, \method{}, a novel method that models personalized diffusive interests and provides suitable recommendations.
    \item \RR{\textbf{Evaluating} the proposed diffusion scheme and recommendation framework through comprehensive experiments. Ablation and parameter studies provide further insights into the functionality of each component.}
\end{itemize}

%% file: Content/2-relatedwork.tex
\section{Related Works}

\subsection{Generative-based Collaborative Filtering}

While the traditional ranking and matching-based methods \cite{huang2023position,wang2023sequential,zhao2022timeaware,liu2023megcf} struggling in modeling the sparse and imbalanced interaction distributions, there are plenty of previous efforts laying their focuses on generative algorithm for making personalized recommendations. Compared with traditional classified-based CF methods \cite{he2020lightgcn,qin2025polycf,zhang2024simplify,wang2023setrank}, generative CF models have the potential to fully exploit the distributional characteristics within interaction systems for comprehensive recommendations.

As a representative generative scheme, Autoencoders \cite{ng2011sparse, baldi2012autoencoders, kingma2013auto}, or AE-based methods have been widely applied in a wide range of recommendation tasks \cite{li2015deep,li2017collaborative,he2018adversarial,wang2022generative}. An AE is typically composed of an optimizable pair of encoder and decoder that cooperate to compress and refactor the data samples, respectively \cite{yu2019understanding, tschannen2018recent}. Before being used for constructing recommender systems, AEs are commonly leveraged for generating images \cite{pu2016variational, lore2017llnet} and texts \cite{li2015hierarchical, yang2017improved}. As a pioneering work in AE-based recommender systems, AutoRec \cite{sedhain2015autorec} uses the similarity of input and output in AE to reconstruct the uninteracted score of the items. CDAE \cite{wu2016collaborative} enhances AutoRec by adding user ids and interacted items as biases on the hidden space to better characterize the bias between different users and items. MultDAE and MultVAE \cite{liang2018variational} turn to variational AEs and interpret the hidden layer as the mean and variance of a series of independent Gaussian distributions. MacridVAE \cite{ma2019learning} disentangles users into prototype clusters to identify user groups. In addition to CF tasks, there are applications of AE models on other recommendation tasks such as event recommendation \cite{zhang2024variational} and user interaction augmentation \cite{xia2021collaborative,wang2022ad}.

\subsection{Diffusion Models}
Recently, diffusion models \cite{wijmans1995solution, rogers2004prospective} have shown the strong functionality on high-quality generation. There are two main paradigms in diffusion models, namely denoising diffusion probabilistic model\cite{ho2020denoising,austin2021structured} and score based generative model\cite{song2020score,anderson1982reverse}. Some researchers have achieved great success in using diffusion models to generate images \cite{ho2020denoising, rombach2022high, kim2022diffusionclip, saharia2022palette}, and it is also widely used in other areas, such as audio synthesis \cite{kong2020diffwave}, text generation \cite{li2022diffusion}, and molecular conformation generation \cite{xu2022geodiff}. CODIGEM \cite{walker2022recommendation} uses diffusion model to generate the recommendation, using only three iterations in forward and reverse processes. DiffRec \cite{wang2023diffusion} considers the inherent noise in input data, and adds more steps in the reverse process. Having realized the limitation of traditional Gaussian-based diffusion on directly generating interaction samples, DDRM \cite{zhao2024denoising} propose to perform diffusion generation process in latent representation space and further utilize the generated user and item representation. There are also applications of diffusion process in downstream tasks such as sequential recommendation models \cite{li2023diffurec,du2023sequential,qin2023diffusion}, federated recommendation \cite{du2025federated}, and multi-modal recomendation \cite{chen2024adversarial}.

%% file: Content/3-preliminary.tex
\section{Preliminary}

\begin{table}[!t]
	\centering
    \caption{Summary of key notations.}
	\begin{tabular}{l  l } 
	\toprule
	  Notation  & Description \\
   \midrule
   $\mathcal{U}$ & The User set. \\
   $\mathcal{I}$ & The Item set. \\
   $R$ & The binary interaction matrix. \\
   $\tilde{R}$ & The normalized interaction matrix. \\
   $\tilde{G}_I$ & The item Gram matrix. \\
   $T$ & Diffusion time range. \\
   $\Delta t$ & Diffusion step size. \\
   $K$ & Granularity of user interests. \\
   $r_u$ & User interaction vector. \\
   $X_u^{(t)}$ & User $u$'s stage-wise interests at diffusion time $t$. \\
   $x_{u,i}^{(t)}$ & User $u$'s stage-wise interests towards a specific item $i$ at $t$. \\
   $\gamma$ & Personalized decay factor weight. \\
   $AE_\theta(\cdot)$ & Autoencoder network parameterized by $\theta$. \\
\bottomrule
\end{tabular}
\label{tab:notation}
\vspace{-0.3cm}
\end{table}
In this section, we will provide a brief overview of graph-based Collaborative Filtering tasks and introduce the relative notations employed in the interests burn-down process to enhance clarity in subsequent sections. 
Since the proposed methods involves multiple notations to describe terminologies of graph-based CF and diffusion process, we summarize the used notations in this paper in Table \ref{tab:notation} for better readability.

\subsection{Graph-based Collaborative Filtering}

Typically, graph-based CF tasks are conceptualized with a user-item bipartite graph that inherently reflects the historical interactions between the user set $\mathcal{U}$ and the item set $\mathcal{I}$. Specifically, the adjacency matrix of the interaction graph $R\in\{0,1\}^{|\mathcal{U}|\times|\mathcal{I}|}$, signifies that $R_{u,i}=1$ indicates the existence of an observed interaction between user $u$ and item $i$. After establishing the adjacency between users and items, a widely applied approach involves normalizing the interaction adjacency with the diagonal degree matrices $D_U$ and $D_I$ that denote the degrees of users and items:
\begin{equation}
    \tilde{R}=D_U^{-\frac{1}{2}}RD_I^{-\frac{1}{2}}.
\end{equation}

For graph-based CF methods, the interaction history of users is presented as rows of the interaction adjacency, denoted as $r_u\in\{0,1\}^{|\mathcal{I}|}$ and the objective of CF models is to reconstruct the missing interactions from $r_u$. 
A common approach to model item-item collaborative correlation is the Gram similarity that obtained from interactions, denoted as $G_I=\tilde{R}^T\tilde{R}$.

\R{The Gram matrix $G_I$ encodes normalized item–item affinities and can support a range of downstream components. While $G_I$ itself has been proven to be a powerful low-pass filter that retrieves similar items from interaction history \cite{shen2021powerful}, it also enables a natural measure of signal smoothness through the Rayleigh quotient:
\begin{equation}
    R_{G_I}(v)=\frac{v^TG_Iv}{v^Tv}=\frac{\sum_{i,j}G_{I_{ij}}v_iv_j}{\sum_{i}v_i^2}.
\end{equation}

Prior studies \cite{liu2023personalized} show that the $R_{G_I}(v)$ characterizes how smoothly a signal $v$ varies over the collaborative similarity structure captured by $G_I$. Higher quotient values typically correspond to more coherent patterns across highly co-consumed items.
}

\subsection{Diffusion Process for Recommendation}

Consider the input space $\mathbb{R}^{|\mathcal{I}|}$ representing potential interactions, the diffusion process consists of both forward and backward processes that depict the transportation from the sample space towards a specific stationary distribution. Typically, a discrete diffusion process is formalized with the transition probability $p(x^{(t)}|x^{(0)})$ and $q(x^{(t-1)}|x^{(t)})$. A score estimation model parameterized by $\theta$ is optimized using the evidence lower bound (ELBO) based on the forward process. During inference, predicted samples are generated following the reversed backward process.

\R{Diffusion models can use various schemes to sample the intermediate samples $x^{(t)}$. In this work, we focus on blackout diffusion, a mechanism designed to capture decay in discrete spaces. Specifically, consider an initial value $x^{(0)}=m\in\mathbb{N}^+$. The diffused state at time $t$ follows:
\begin{equation}
x^{(t)}\sim p(x^{(t)}|x^{(0)}=n)=\sum_{k\le n}p(x^{(t)}=k|x^{(0)}=n),
\end{equation}
where the conditional distribution of $x^{(t)}$ is modeled as a binomial decay. The integers $k,n$ indicate the discrete value of the sampled $x^{(t)}$ and original $x^{(0)}$. Under this scheme, each unit of the initial value independently "survives" with probability $e^{-t}$. Thus, $x^{(t)}$ is sampled from:
\begin{equation}
    x^{(t)}\sim Binomial(x^{(0)},e^{-t}),
\end{equation}
with probability mass function of
\begin{equation}
    p(x^{(t)}=k|x^{(0)}=n)=\tbinom{n}{k}e^{-kt}(1-e^{-t})^{(n-k)},\ \tbinom{n}{k}=\frac{n!}{k!(n-k)!}.
\end{equation}

This binomial decay guarantees a simple Markov process in which the state monotonically decreases over time. Because the survival probability diminishes exponentially, the process converges at a controlled rate to its stationary distribution:
\begin{equation}
    \label{eq:general_transition}
    \lim\limits_{t\rightarrow\infty}x^{(t)}=\boldsymbol{0}.
\end{equation}
}


To make personalized recommendations based on the obtained score estimation model, the inference phase starts with a initialized $\hat{x}^{(T)}$ that is drawn directly from the user's interaction $r_u$, and is iteratively updated via the learned denoising model to produce the recommendation result.


%% file: Content/4-methodology.tex
\section{Methodology}

\begin{figure*}
\centering
\includegraphics[width=\linewidth]{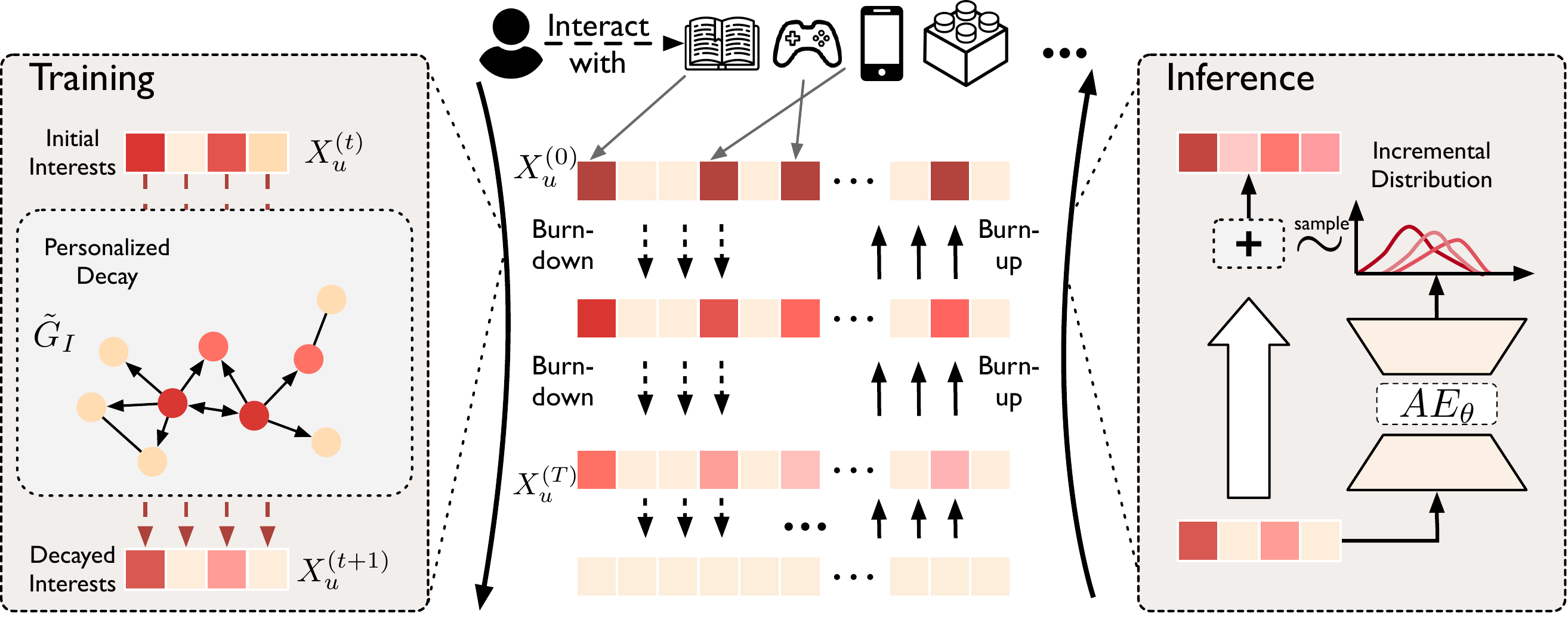}
\caption{The overall illustration of \method{} and interests burn-down process.} 
\label{fig:main}
\end{figure*}

In this section, we will introduce the proposed interests burn-down process in collaborative filtering tasks, accompanied by the training and inference algorithm of the score estimation model. The interests burn-down process is designed for depicting the stage-wise interests vanishing process of users in interaction systems. A graph convolution kernel is employed to introduce collaborative similarity between items, contributing to a personalized burn-down process. Furthermore, the score estimation model undergoes optimization through the interests burn-down process, and the recommendation results are derived through the reverse burn-up process. A comprehensive illustration of the proposed framework \method{} is presented in Figure \ref{fig:main}.

\subsection{Interests Burn-down Process}

To exploit the potential in the diffusion process of users' interaction interests, we introduce the personalized interests burn-down process to model the evolution of preferences in interaction systems. In contrast to existing methods that directly employ Gaussian diffusion noises to user interactions, the proposed interests burn-down process is specifically tailored to seamlessly adapt and simulate the real-world evolution process of users' interests.

\subsubsection{Burn-down Forward Process}
Considering a user's interests in each item within a given interaction system, over time, it is natural for the user's interest to eventually diminish at different levels. The interest in certain items may also trigger interest in other items during this process. Motivated by this observation, we propose an interests burn-down process that illustrates how a user's interests in items gradually fade away independently. By modeling it this way, the interests of all users would finally decay to zero, forming a stationary distribution.

As mentioned earlier, the user's interaction history $r_u$ is modeled as binary vectors in collaborative filtering. To capture the finely-grained interests evolution process, we extend the interaction vectors to a multi-stage style and obtain the stage-wise interests $X_u=Kr_u$, where $K\in\mathbb{N}^+$ is a predefined hyperparameter that defines the granularity of user interests.

With the stage-wise user interests $X_u$ defined, we present the proposed interests burn-down process. \R{To extend the idea of blackout diffusion into interest space, we first propose a generalized burn-down process for interests vectors. Specifically, the forward transition process given arbitrary diffusion time $t$ can be regarded as sampling from the binomial decay distribution:
\begin{equation}
\label{eq:forward}
    X_u^{(t)}\sim Binomial(X_u^{(0)},e^{-t}).
\end{equation}

This transition imposes an independent exponential decay on each component of the interest vector. Although the process converges to an all-zero vector in the limit, the intermediate state $X_{u}^{(t)}\in\mathbb{N}^{|\mathcal{I}|}$ still retains meaningful information about the user's original preferences, enabling effective reconstruction.
}

\subsubsection{Personalized Graph Convolution}

\R{The transition probability of vanilla blackout diffusion in Equation \ref{eq:general_transition} provides a natural backbone for modeling the discrete interests burn-down behavior. However, prior empirical work shows that users’ preferences do not decay uniformly. Specifically, interests tied more strongly to a user’s behavioral profile tend to persist longer, while weaker or peripheral interests decay more quickly. To capture these heterogeneous patterns, we incorporate collaborative information directly into the diffusion dynamics. We achieve this by introducing a graph-convolution-based decay factor that modulates the binomial decay rate for each item. The enhanced forward process becomes:} 


\begin{equation}
\label{eq:forward_decay}
X_u^{(t)}\sim Binomial(X_u^{(0)},e^{-\boldsymbol{F}_u(t)}),
\end{equation}
where $\boldsymbol{F}_u(t)\in\mathbb{R}^{|\mathcal{I}|}$ represents the personalized interests decay factor that controls the burn-down efficiency of a user's interests accordingly.

To introduce collaborative relationships between users and items to the decay factor $\boldsymbol{F}_u(t)$, we draw inspiration from previous researches on graph signal processing \cite{shen2021powerful,qin2025polycf} and apply a graph convolution operator to incorporate personalized collaborative signals. Utilizing the normalized interaction matrix $\tilde{R}$, the interests decay factor is defined as:
\begin{equation}
\label{eq:decay_factor}
[\boldsymbol{F}_u(t)]_{i}=\frac{t}{1+[\gamma(\tilde{R}^T\tilde{R})r_u]_i}=\frac{t}{1+[\gamma\tilde{G}_Ir_u]_i},
\end{equation}
where $\gamma$ is a predefined hyperparameter controlling the impact of item-level collaborative effect. The Gram matrix $\tilde{G}$ serves as a graph filter to extract more informative and crucial low-pass components within interactions. 
The introduced collaborative decay factor encourages the preservation of user interests towards collaboratively related items, thereby maintaining the personalized user portraits would remain throughout interests burn-down process.

\begin{figure}
\centering
\includegraphics[width=0.95\linewidth]{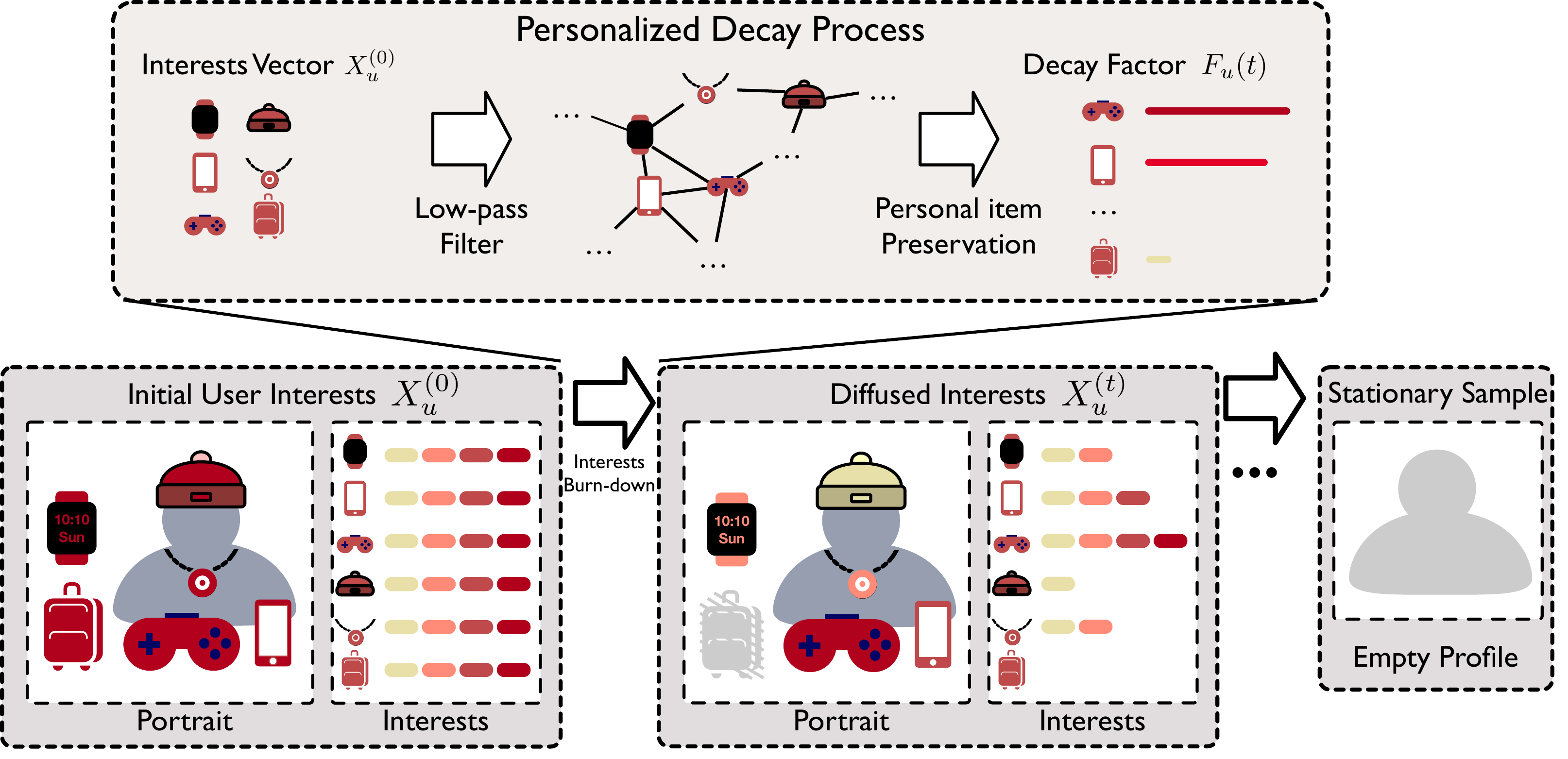}
\caption{The illustration of the personalized interests decay process, where the user portraits are described as discrete interest density towards different items. Throughout the personalized decay, a low-pass item graph filter is applied to obtain personalized decay factor $\boldsymbol{F}_u(t)$ for each item and steers the decay probability of each interest dimension. The decay factor helps preserve the interactive characteristics of user portraits.}
\label{fig:decay}
\end{figure}

As illustrated in toy example from Figure \ref{fig:decay},  the proposed decay factor $\boldsymbol{F}_u(t)$ acts as a personalized damping term, selectively slowing down the decay of user interests in items that are more aligned with their historical behavior.
For instance, in the shown user profile, which reveals a preference for electronic devices and fashion, $\boldsymbol{F}_u(t)$ dynamically steers the decay process such that relevant items (e.g., e-watch and controller) fade slower than unrelated ones (e.g., suitcases). This selective decay is achieved by combining exponential dynamics with collaborative graph filtering, resulting in the following two key properties:
\begin{itemize}
    \item At both the item and spectral levels, components that are more aligned with the user's past interactions exhibit slower decay rates.
    \item The stationary distribution of the personalized decay process remain zero state.
\end{itemize}

\textbf{The personalized characteristics of the decay dynamics.} To characterize the personalized decay dynamics, we continuous-approximate the binomial sampling process via differential analysis. 

The expected value of the diffused sample follows:
\begin{equation}
    \mathbb{E}[x_{u,i}^{(t)}]=x_{u,i}^{(0)}e^{-\frac{t}{1+[\gamma\tilde{G}_Ir_u]_i}}.
\end{equation}

\R{By taking the time derivative of the expectation, we can obtain the expected decay rate of each interest component evloves during forward diffusion:
\begin{equation}
    \frac{\mathrm{d}}{\mathrm{dt}}\mathbb{E}[X_{u,i}^{(t)}]=-\frac{1}{I+[\gamma\mathrm{diag}(\tilde{G}_Ir_u)]_i}\mathbb{E}[X_{u,i}^{(t)}].
\end{equation}

Stacking all item dimensions gives an approximation on the dynamics of $X_u^{(t)}$ with the time derivative of its expectation:
\begin{equation}
    \frac{\mathrm{d}}{\mathrm{dt}}X_{u}^{(t)}=-(I+\gamma\mathrm{diag}(\tilde{G}_Ir_u))^{-1}X_u^{(t)}.
\end{equation}
}

\R{This defines a linear ODE that describes the pure-death process, governed by a damping matrix} $M=(I+\gamma\mathrm{diag}(\tilde{G}_Ir_u))^{-1}$, which encodes item-specific decay rates based on collaborative similarity. Consider the spectral decomposition $\tilde{G}_I=V\Lambda V^{T}$, where $V=[v_1,...,v_n]\in\mathbb{R}^{n\times n}$ are orthonormal eigenvectors, and the $\Lambda=\mathrm{diag}(\lambda_1,...,\lambda_n)$ contains eigenvalues.
\R{We quantify the spectral similarity between a basis $v$ and interaction $r_u$ using the Rayleigh quotient on a user-induced affinity matrix:
\begin{equation}
    R(v)=v^T\mathrm{diag}(\tilde{G}_Ir_u)v=\sum_{i=1}^nv_i^2[\tilde{G}_Ir_u]_i.
\end{equation}
Since each $v$ is normalized, the denominator of the standard Rayleigh quotient is omitted.
}

\R{Intuitively, $[\tilde{G}_Ir_u]_i$ reflects how strongly item $i$ is connected to user $u$'s historical profile. Hence the Rayleigh quotient $R(v)$ measures the extent to which the eigenvector $v$ concentrates on user-relevant items. For the diffused interests vector $X_u^{(t)}$, it admits a spectral decomposition:
\begin{equation}
X_u^{(t)}=\sum\limits_{i=1}^n\left<X_{u}^{(t)},v_i\right> \cdot v_i=x_{u,v}\cdot v_i,
\end{equation}
where $\left<\cdot,\cdot\right>$ denotes the inner product. For each eigen vector $v_i$, the coefficient $x_{u,v}=\left<X_{u}^{(t)},v_i\right>$ characterizes the contribution of that spectral component, and its temporal evolution becomes:
\begin{equation}
    \frac{\mathrm{d}}{\mathrm{dt}}x_{u,v}^{(t)}=-v^TMvx_{u,v}^{(t)}=-\frac{1}{1+\gamma R(v)}x_{u,v} 
\end{equation}
}

Compared with the item-wise interests density, the projected spectral density better reflects the collaborative characteristics of user's historical behavior. Consider the personalized decay on the specific spectral direction $v$, when comparing the the decay rate on the item dimension $i$, we have:
\begin{equation}
\label{eq:spectral_decay}
\begin{cases}
    \frac{\mathrm{d}}{\mathrm{dt}}x_{u,v}^{(t)}=-\frac{1}{1+\gamma R(v)}x_{u,v} \\
    \frac{\mathrm{d}}{\mathrm{dt}}x_{u,i}^{(t)}=-M_ix_{u,i}^{(t)}=-\frac{1}{1+\gamma [\tilde{G}_Ir_u]_i}x_{u,v}
\end{cases}
\end{equation}

From Equation \ref{eq:spectral_decay} we can conclude that:
\begin{itemize}
    \item From {spectral level}, for spectral components $v,w$ of $\tilde{G}_I$, the interests burn down on $v$  is slower than $w$ if and only if $R(v)>R(w)$, i.e.,  $v$ aligns more with $r_u$ in the spectral space.
    \item From {item level}, for any item $i,j$ that have interacted with $u$, the decay on item $i$ is slower than on $j$ if and only if $[\tilde{G}_Ir_u]_i>[\tilde{G}_Ir_u]_j$, i.e., $i$ more collaborative-relevant to user $r_u$.
\end{itemize}

\textbf{The stationary distribution}
Consider the item $i$ of $X_u^{(t)}$ during the interests burn-down, since the $\gamma[\tilde{G}_Ir_u]_i$ part in Equation \ref{eq:decay_factor} remain constant over time for each item, we have the remain probability converges to 0:
\begin{equation}
    \lim_{t\rightarrow\infty}e^{-\boldsymbol{F}_u(t)}=\lim_{t\rightarrow\infty}e^{-\frac{t}{1+[\gamma\tilde{G}_Ir_u]_i}}=0.
\end{equation}

Furthermore, the survival probability under the binomial sampling process satisfies:
\begin{equation}
    Pr[x_{u,i}^{(t)}>0]=1-(1-e^{-\boldsymbol{F}_u(t)})^{x_{u,i}^{(0)}}\le e^{-\boldsymbol{F}_u(t)}\cdot x_{u,i}^{(0)}\rightarrow 0\ \ (t\rightarrow\infty).
\end{equation}

Thus, each item component vanishes in probability as $t$ approaches infinity, ensuring that the stationary distribution is the zero vector. In other words, the diffused samples satisfy:
\begin{equation}
    x_{u,i}^{(t)}\stackrel{p}{\longrightarrow}0,\ \forall i\in \mathcal{I}
\end{equation}

\subsection{Reversed Burn-up Process}

In the burn-down process of user interests, a corresponding reverse-time sampling process is employed to model the generative mechanism, whereby user interests burn-up towards various items, thereby formulating personalized recommendation results.


\subsubsection{Burn-up Backward Process}
As presented in the forward burn-down process in Equation \ref{eq:general_transition}, the conditional distribution of arbitrary $p(x_{u,i}^{(t)}|x_{u,i}^{(0)})$ is fully determined by the initial count $x_{u,i}^{(0)}$. \R{For reverse diffusion, however, we require a tractable expression for sampling from the backward transition $p(x_{u,i}^{(t-\Delta t)}|x_{u,i}^{(t)})$ at each time step. To derive this, we first compute the forward adjacent transition over a small interval $\Delta t$} by substituting $x_{u,i}^{(0)}$ with $x_{u,i}^{(t-\Delta t)}$:
\begin{equation}
\label{eq:step_forward}
    p(x_{u,i}^{(t)}=k|x_{u,i}^{(t-\Delta t)}=m)=\tbinom{m}{k}e^{-k\Delta t}(1-e^{-\Delta t})^{m-k}.
\end{equation}

\R{With this forward step in hand, the corresponding backward transition can be expressed using Bayes’ rule:}
\begin{equation}
    p(x_{u,i}^{(t-\Delta t)}|x_{u,i}^{(t)},x_{u,i}^{(0)})=\frac{p(x_{u,i}^{(t)}|x_{u,i}^{(t-\Delta t)})p(x_{u,i}^{(t-\Delta t)}|x_{u,i}^{(0)})}{p(x_{u,i}^{(t)}|x_{u,i}^{(0)})}.
\end{equation}

With the substitution of the probability value from Equation \ref{eq:general_transition} and \ref{eq:step_forward}, we have the conditional probability represented as:
\begin{align}
\label{eq:reverse_prob}
    p(x_{u,i}^{(t-\Delta t)}=m|x_{u,i}^{(t)}=k,x_{u,i}^{(0)}=n)
    =&\frac{p(x_{u,i}^{(t)}=m|x_{u,i}^{(t-\Delta t)}=k)p(x_{u,i}^{(t-\Delta t)}=k|x_{u,i}^{(0)}=n)}{p(x_{u,i}^{(t)}=m|x_{u,i}^{(0)}=n)} \\
    =&\frac{\tbinom{m}{k}\cdot\tbinom{n}{m}}{\tbinom{n}{k}}\cdot\frac{e^{-k\Delta t}(1-e^{-\Delta t})^{m-k}e^{-m(t-\Delta t)}(1-e^{-(t-\Delta t)})^{n-m}}{e^{-m(t-\Delta t)}(1-e^{-(t-\Delta t)})^{n-m}} \\
    =&\frac{(n-k)!}{(m-k)!(n-m)!}(\frac{1-e^{-(t-\Delta t)}}{1-e^{-t}})^{n-m}(\frac{e^{-(t-\Delta t)}-e^{-t}}{1-e^{-t}})^{m-k}.
\end{align}

So far, by introducing the notation of:
\begin{equation}
    r_t=\frac{e^{-(t-\Delta t)}-e^{-t}}{1-e^{-t}},
\end{equation}

we are able to simplify Equation \ref{eq:reverse_prob} and reformulate the reverse transition rate as:
\begin{align}
\label{eq:backward}
    p(x_{u,i}^{(t-\Delta t)}=m|x_{u,i}^{(t)}=k,x_{u,i}^{(0)}=n)&=\frac{(n-k)!}{(m-k)!(n-m)!}r_t^{n-m}(1-r_t)^{n-m} \\
    &=\tbinom{n-k}{m-k}r_t^{n-m}(1-r_t)^{n-m}.
\end{align}

\subsubsection{Making Generative Recommendations}

In practice, the recommended interests vector $X_{u}^{(0)}$ is obtained with a neural network model to iteratively update $X_u^{(t)}$. In our burn-up process case, we adopt the Binomial bridge formula \cite{santos2023blackout} and construct the generative process with the observation of:
\begin{equation}
\label{eq:backward_distribution}
    X_{u}^{(t-1)}-X_u^{(t)}\sim Binomial(X_u^{(0)}-X_u^{(t)},r_t).
\end{equation}

From Equation \ref{eq:backward_distribution} we observe that the reverse generative process can be computed through iterative sampling from specific Bernoulli processes. Consequently, we implement the estimation of the time of Bernoulli trials $X_u^{(0)}-X_u^{(t)}$ with an autoencoder-based estimation model $AE_\theta(X_u^{(t)},t)$ parameterized by $\theta$:
\begin{equation}
    q_u(t)=\mathrm{softplus}(AE_\theta(X_u^{(t)},t))=\log(1+e^{AE_\theta(X_u^{(t)},t)})
\end{equation}

With the estimated Bernoulli trail parameters, the recommendation burn-up process is subsequently transformed into an inference-sample loop via the estimation model.  In the context of our \method{}, we construct $AE_\theta(\cdot)$ with a multi-layer perception (MLP) aligning with previous diffusion-based CF research \cite{wang2023diffusion}. While other specifically designed model structures may be more applicable to represent user's interaction behaviour, we aim to demonstrate that, under the assumption of the personalized interests burn-down process, simple structures like MLPs are capable enough of achieving promising recommendation performance.

\subsection{Model Optimization and Inference}

\subsubsection{Model Optimization}

\input{Tables/algorithm2}
As mentioned in previous sections, \method{} generates the personalized recommendation results through sampling from a series of Bernoulli distributions, guided by an estimation network $q_u(t)$. To ensure alignment between the model-predicted distribution and real-world interaction scenarios, an objective function to match the distributions is required. Consequently, the score estimation network is optimized by minimizing the KL-divergence between the predicted Bernoulli distribution and the ground-truth burn-down distribution. Given the ground-truth burn-down rate in the forward process:
\begin{equation}
    \lambda(t)=\lim_{\delta t\rightarrow 0}p(X_u^{(t+\delta t)}|X_u^{(t)},X_u^{(0)}),
\end{equation}
the optimization target is derived as follows:
\begin{align}
\label{eq:loss_kld}
    \mathcal{L}&=D_{KL}(\lambda(t)\delta t\ \Vert\ \kappa_u(t)\delta t)\\
    &=\lambda(t)\delta t\log\frac{\lambda(t)\delta t}{\kappa_u(t)\delta t}+(1-\lambda(t)\delta t)\log\frac{1-\lambda(t)\delta t}{1 - \kappa_u(t)\delta t}.
\end{align}
Where $\kappa_u(t)$ represents the predicted burn-down probability for user $u$ at arbitrary time step $t$. With the $\theta$-dependent output $q_u(t)$ serving as an approximation of transitions $X_u^{(0)}-X_u^{(t)}$, we have:
\begin{equation}
    \kappa_u(t)=q_u(t)\cdot \frac{e^{-\boldsymbol{F}_u(t)}}{1-e^{-\boldsymbol{F}_u(t)}}.
\end{equation}

After neglecting the terms irrelative to $\theta$, the divergence term in Equation \ref{eq:loss_kld} would be expanded as:
\begin{equation}
\label{eq:delta_loss}
    \mathcal{L}=\kappa_u(t)-\lambda(t)\log\kappa_u(t).
\end{equation}

Note that the objective in Equation \ref{eq:delta_loss} only represents the divergence in a short period of time $\Delta t$. To optimize the model at the whole time period, we need to apply the Monte Carlo sampling methods to all possible $t$. For the binomial-decayed forward process, the $\lambda(t)$ can be obtained with:
\begin{equation}
    \lambda(t)=(X_u^{(0)}-X_u^{(t)})\frac{e^{-\boldsymbol{F}_u(t)}}{1-e^{-\boldsymbol{F}_u(t)}}.
\end{equation}
Recall that $\kappa_u(t)=q_u(t)\frac{e^{-\boldsymbol{F}_u(t)}}{1-e^{-\boldsymbol{F}_u(t)}}$, we can further rewrite the loss function of \method{}:
\begin{align}
\label{eq:loss_fn}
    \mathcal{L}_{\method{}}&=\kappa_u(t)-(X_u^{(0)}-X_u^{(t)})\frac{e^{-\boldsymbol{F}_u(t)}}{1-e^{-\boldsymbol{F}_u(t)}}\log\kappa_u(t)\\
    &=e^{-\boldsymbol{F}_u(t)}(q_u(t)-(X_u^{(0)}-X_u^{(t)})\log q_u(t))+C,
\end{align}
after ignoring the $\theta$-independent terms $C$ and rescale the objective with constant.

From the perspective of uniformed diffusion generative models \cite{ho2020denoising}, the loss function of \method{} in Equation \ref{eq:loss_fn} can be interpreted as the evidence lower bound (ELBO) with an importance sampling weight \cite{nichol2021improved} of $e^{-\boldsymbol{F}_u(t)}$, placing more emphasis on the early stages of the interests burn-down process for better optimization. 

To make an illustrative view of the optimization algorithm of \method{}, we present the training algorithm in Algorithm \ref{alg:2}.

\subsubsection{Burn-up Recommendation}

In the preceding sections, we have introduced the reverse process of the interests burn-up process. To generate personalized recommendation results for users, we employ the interests burn-up process outlined in Equation \ref{eq:backward_distribution}. The reverse process is utilized to sample the incremental interests vector at each time step. Specifically, we iteratively sample the delta of interests $X_u^{(t)}$ from $Binomial(q_u(t),r_t)$ to obtain the updated $X_u^{(t-1)}$ from the initial sample $X^{(T')}$, where we set $X^{(T')}=Kr_u$ by default. Finally, the candidate items with the largest interests scores are selected for the downstream top-K recommendation and evaluation. The overall recommendation procedure of \method{} is summarized in Algorithm \ref{alg:1}.

\input{Tables/algorithm1}

Since the incremental nature of the burn-up process, the recommendation score of each candidate item will eventually reach the maximum of $K$ as the reverse time $T'$ increases to infinity, which underscores the significance of choosing an appropriate $T'$ for achieving optimal recommendation performance.

\subsection{Computational Complexity of Interests Burn-down Process}
In real-world applications, computational costs are a pivotal consideration. Consequently, we present the computational costs associated with the proposed interests burn-down process and the regular Gaussian diffusion-based models.

Consider an interaction system with $|\mathcal{U}|$ users, $|\mathcal{I}|$ items, and $r$ interactions. To optimize a diffusion-based model with $s$ time steps, a regular Gaussian-Markov process requires the iterate of forward diffusion steps on each items in the interaction vector, leading to a complexity of $\mathcal{O}(s|\mathcal{U}||\mathcal{I}|)$. On the other hands, the interests burn-down process only needs to focus on the interacted items, reducing the complexity into $\mathcal{O}(sr)$. Introduction of the personalized decay factor will increase the computational costs, while with the fully leveraged sparsity of interactions. The complexity with personalized convolution is $\mathcal{O}(s\frac{r^2}{|\mathcal{U}|})$, which is still lower than Gaussian diffusion, given the fact that in sparse interacting systems, the interaction $\mathcal{O}(r)=\mathcal{O}(\mathcal{U}+\mathcal{I})$.

During the inference stage, both methods must consider all the candidate items to generate recommendations. To iteratively sample the recommendation results for all users, both Gaussian diffusion models and interests burn-down models share the complexity of $\mathcal{O}(s|\mathcal{U}||\mathcal{I}|)$, ignoring the complexity brought by the autoencoder network.

%% file: Tables/algorithm2.tex
\begin{algorithm2e}
\LinesNumberedHidden
\SetNlSkip{0em}
\SetAlgoNlRelativeSize{0}
\caption{Training Procedure of \method{}}
\label{alg:2}
\KwIn{Interaction set $R$; Sample step $T$}
\SetKwInput{kwInit}{Initialize}
\kwInit{Gram matrix $\tilde{G}_I$; Model parameters $\theta$}
\While{not converged}{
Sample interaction $R_u$ from $R$\;
Initialize user interests $X_u^{(0)}\leftarrow Kr_u$\;
Draw time $t\in\mathbb{R}$ from $[1,T]$\;
Obtain personalized decay $\boldsymbol{F}_u(t)\leftarrow \frac{t}{1+\gamma\tilde{G}_Ir_u}$\;
Perturbed $X_u^{(t)}\sim Binomial(X_u^{(0)},e^{-F_u(t)})$ \;
Optimize $\theta$ with $\mathcal{L}_{StageCF}$ in Equation \ref{eq:loss_fn}\;
}
\Return{$\theta$\;}
\end{algorithm2e}

%% file: Tables/algorithm1.tex
\begin{algorithm2e}[t]
\LinesNumberedHidden
\newcounter{algoline}
\newcommand\Numberline{\refstepcounter{algoline}\nlset{\thealgoline}}
\SetNlSkip{0em}
\SetAlgoNlRelativeSize{0}
\caption{Recommendation Procedure of \method{}}
\label{alg:1}
\KwIn{Model parameter $\theta$, observed user interactions $r_u$}
Initialize user interests $X_u^{(T')}\leftarrow Kr_u$\;
\For{$t=T'$\ to\ $1$}{
$q\_logits\leftarrow \mathrm{softplus(AE_\theta(X_u^{(t)},t))}$ \;
$q_u(t)\leftarrow$ Clamp$(q\_logits,0,K-X_u^{(t)})$\;
$\Delta X_u\sim Binomial(q_u(t),r_t)$\;
$X_u^{(t-1)}\leftarrow X_u^{(t)}+\Delta X_u$\;
}
Recommend Set $R_u\leftarrow\mathrm{TopK}(X_u^{(0)})$\;
\Return{$R_u$\;}
\end{algorithm2e}

%% file: Content/5-experiment.tex
\section{Experiment}

This section presents the comprehensive experiments on collaborative filtering task. We evaluate \method{} alongside representative baseline methods using three widely-used interaction datasets to demonstrate its superiority and effectiveness. Additionally, we conduct in-depth ablation and parameter studies to investigate the functionality of the proposed interest burn-down diffusion process. The implementation of \method{} and experimental code is publicly available at \url{https://github.com/Yifang-Qin/StageCF}.

\subsection{Experimental Setup}

\subsubsection{Datasets and Evaluation Metrics}

We evaluate the model's performance using three widely-applied user-item interaction datasets: \textbf{Gowalla}, \textbf{Yelp2018}, and \textbf{Amazon-Book}. Table \ref{tab:statics} presents the statistics of these datasets. Specifically, we utilize the train/test splits and data settings provided by previous work, LightGCN \cite{he2020lightgcn}\R{, where the 80\% of the interactions are selected as training and valiadation set, while the remaining as test set.} \RR{Specifically, we randomly draw 10\% of training interactions as the validation set to select the optimal training parameters and the training epochs.}

For all datasets mentioned, we employ two widely accepted CF benchmark metrics: Recall@K and NDCG@K. Here, the K is chosen from 10, 20, and 50 to evaluate model performance across different scales for making comprehensive comparison. Recall@20 on the validation set is selected as the criterion for both hyperparameter selection and the number of training epochs. Hyperparameter search and final evaluation follow a two-stage protocol. (1) During search, each configuration is trained while monitoring validation Recall@20 every epoch; for the iterative methods (\method{}, LightGCN, MultVAE, MultDAE, MacridVAE, DiffRec, DDRM, and FlowCF) we adopt early stopping with patience of 5 epochs and record the epoch at which the highest validation Recall@20 is attained. SLIM and iALS are non-iterative---SLIM is fitted to its convergence tolerance via sklearn ElasticNet, while iALS uses a fixed schedule of 15 ALS iterations---so early stopping does not apply. (2) For final evaluation, each method is retrained from scratch with the selected hyperparameters and, for the iterative methods, for exactly the number of epochs that achieved the best validation Recall@20; the resulting model---corresponding to the best-validation-epoch checkpoint reproduced via deterministic retraining---is evaluated once on the held-out test set, which is never used during search.
\input{Tables/statistics}

\subsubsection{Baseline Models}

To demonstrate the superiority of the proposed method, we conduct performance comparison between \method{} and a diverse set of baseline CF methods that includes:

\begin{itemize}[leftmargin=*]
    \item \textbf{Graph-based CF models} that include \R{SLIM \cite{xia2011slim}, \RR{iALS} \cite{hu2008collaborative} and LightGCN \cite{he2020lightgcn}}. These methods leverage the adjacency graph of user-item interactions to make recommendations.
    \item \textbf{Autoencoder-based CF models} that include MultDAE/MultVAE \cite{liang2018variational}, and MacridVAE \cite{ma2019learning}. These methods focus on constructing encoder-decoder architectures to rank candidate items through interaction vectors.
    \item \textbf{Diffusion-based CF models} that include 
    DiffRec \cite{wang2023diffusion}, DDRM \cite{zhao2024denoising}, \R{and FlowCF \cite{liu2025flowcf}}. These pioneering works explore application of different diffusion processes in recommendation tasks.
\end{itemize}


{We implement the aforementioned baseline methods based on their open-source code or their original papers. Both \method{} and the baselines are developed useing PyTorch.}
\RR{For all the embedding-based method like LightGCN and DDRM, the embedding size is tuned from $\{64,128,256,512,1024\}$. For the rest autoencoder-based method, the hidden sizes of the encoder MLPs are tuned from $\{[600,200],[500,300],[1000]\}$ and the decoders take the reverse hidden sizes. The layer of LightGCN is tuned from $\{1,2,3\}$. We adapt the same setting of SGL backbone in DDRM~\cite{zhao2024denoising}.}

\RR{Regarding training details, the learning rate for all models are tuned from $\{5\times10^{-5},10^{-4},10^{-3}\}$ with the dropout ratio tuned from $\{0.3,0.4,0.5\}$. Specifically, for \method{}, the maximum stage num $K$ is searched from $\{200,250,300,350,400\}$, the forward diffusion time are fixed at $T=T'=4$, and the sampling time step number is set to $100$ by default. For all the aforementioned models, the training batch size is tuned from $\{512, 1024, 2048\}$. We fine-tune the set of hyper-parameters with grid search method and select the best performance of Recall@20 on validation set to represent the model's overall performance. The per-method search ranges follow the tuning recommendations from each method's original publication and related literature on hyperparameter sensitivity, so the search budgets across methods are comparable in scope rather than identical in cardinality. While previous researches have proposed the sensitivity of linear methods towards hyper-parameters~\cite{rendle2022revisiting,ferrari2019we}, we follow their detailed advice to fully tune the SLIM and iALS methods to ensure a more fair comparison. All hyperparameters are jointly tuned via grid search.
}

\subsection{Performance Comparison}

\input{Tables/main_result}
We conduct general experiments to evaluate the recommendation performance of \method{} and the aforementioned baseline models, and the results are presented in Table \ref{tab:overall}. From the results we can observe that:
\begin{itemize}[leftmargin=*]
    \item The proposed \method{} surpasses the baseline methods across all genres, underscoring its superiority over conventional diffusion and autoencoder-based recommendation schemes. Notably, the \method{} exhibits superior performance compared to the most competitive baseline methods by over 4.7\%, 5.5\%, and 14.2\% on Recall@20, and by 4.3\%, 4.4\%, and 11.3\% on NDCG@20 across the three datasets.
    \item The diffusion process demonstrates significant potential in modeling collaborative interaction systems. In comparison with most autoencoder-based methods, the assumption of diffusive user interaction behavior enables the model to achieve more promising recommendation performance.
    Whether employing a denoising reconstruction scheme (DiffRec) or a diffusion generative scheme (FlowCF and \method{}), the experimental results underscore the capability of the diffusion process in depicting flexible and personalized user interests.
    \item Applying the Gaussian diffusion process directly to interaction vectors (DDRM and DiffRec) yields sub-optimal results due to a mismatch between the Gaussian diffusion process and the nature of user interaction behavior. \method{} maximizes the expressiveness of diffusion models by introducing a tailor
    -designed personalized interests burn-down process.
\end{itemize}

\subsection{Ablation Study}

The proposed \method{} relies on a personalized interests burn-down and its corresponding reversed burn-up process. 
\RR{Compared with Gaussian diffusion that applies uniform corruption to all interactions, the personalized interests burn-down process introduces a structured decay mechanism over user-item interactions. This design implicitly controls how different items vanish during the forward process, especially preserving low-frequency interactions for longer steps. As a result, these items are more likely to be recovered in the reverse process, which is critical for improving recommendation performance.}
To gain a deeper understanding of the functionality of the proposed diffusion process and how \method{} benefits from the personalized interests decay, we conduct comprehensive ablation studies to unveil the role and impact of each element in \method{}.

\subsubsection{Functionality of Diffusion Process}
The interests burn-down diffusion process is grounded in the stage-wise interests vector, as well as the decay process from multiple binomial distributions. To assess the advantage of the proposed process, we conduct ablation studies, comparing the interests burn-down process and Gaussian diffusion process that have been adapted in CODIGEM and DiffRec. Specifically, we compare these two diffusion schemes by conducting experiments involving:
\begin{itemize}[leftmargin=*]
    \item \RR{\textbf{M0 (without diffusion)}}: Removing the entire diffusion process and relying solely on the network structure to make recommendations, akin to the AE-based baseline models.
    \item \RR{\textbf{M1 (one-step sampling)}}: \RR{Training the model with normal diffusion objective, while replacing the sampling process with a one-step output from the filter model. This variant isolates the filtering component without iterative diffusion.}
    \item \textbf{M2 (denoising style)}: Reforming the training and inference procedure with a denoising process, optimizing the model with the MES reconstruction loss. It is also denoted as $x_0$-ELBO as descriped in DiffRec paper \cite{wang2023diffusion}.
    \item \textbf{M3 (diffusive style)}: Applying the diffusion generative process to directly produce recommendation results via backward process, optimizing the model through the ELBO target function. It is also denoted as $\epsilon$-ELBO as descriped in DiffRec.
\end{itemize}

\input{Tables/ablation_diffusion}

By comparing the results of these variants, we gain more
insights into the characteristics of different diffusion scheme in recommender systems. Observing the results in Table \ref{tab:abla_diff}, we note that:
\begin{itemize}[leftmargin=*]
    \item The diffusion processes significantly enhance the expressiveness of existing network structures in making more appropriate recommendations. \RR{Compared with model without any forms of diffusion (the M0 variant) and model without diffusion-based sampling (the M1 variant), both the diffusive styles of training and sampling processes significantly improve model performance.}
    \item Compared with Gaussian diffusion process, the proposed interests burn-down process improves the model in both diffusion styles. This illustrates that the burn-down process is more suitable for CF tasks and can depict the decay dynamics of user's interests.
    \item For interests burn-down process, the diffusion generative scheme can fully exploit its potential in making personalized recommendations. On the other hands, the Gaussian diffusion only functions well when employing the denoising diffusion scheme.
\end{itemize}

\subsubsection{Effectiveness of Personalized Interests Burn-down}

\begin{figure}
\centering

\begin{subfigure}{0.32\linewidth}
    \includegraphics[width=\linewidth]{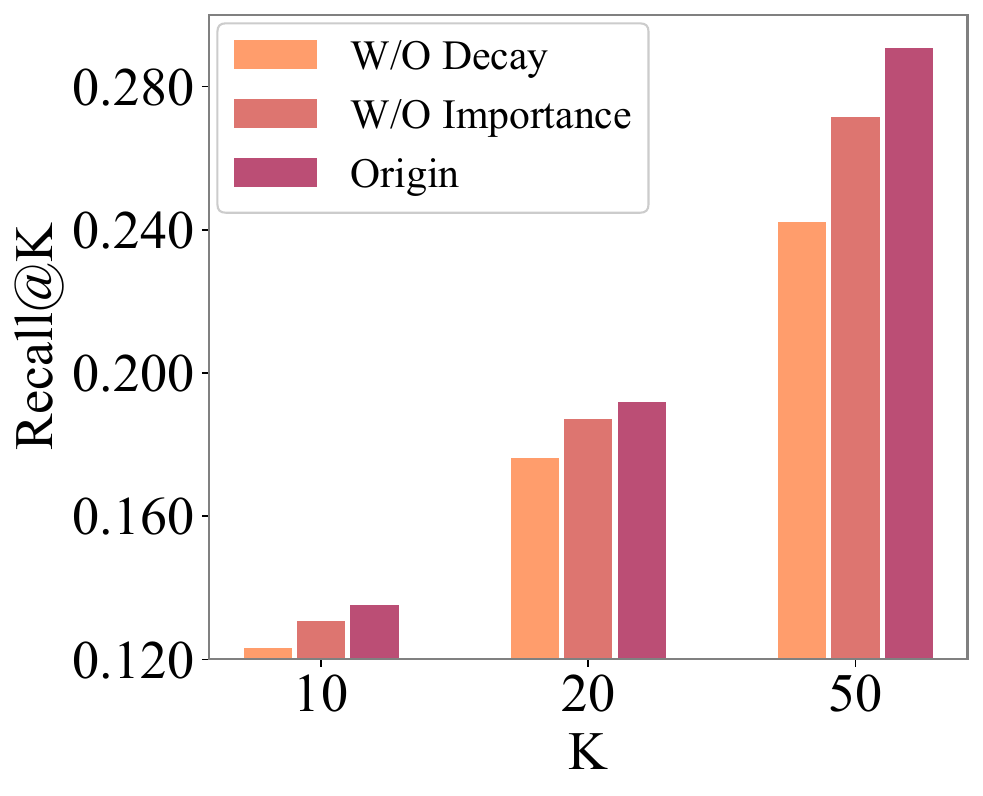}
    \caption{Recall on Gowalla}
\end{subfigure}
\begin{subfigure}{0.32\linewidth}
    \includegraphics[width=\linewidth]{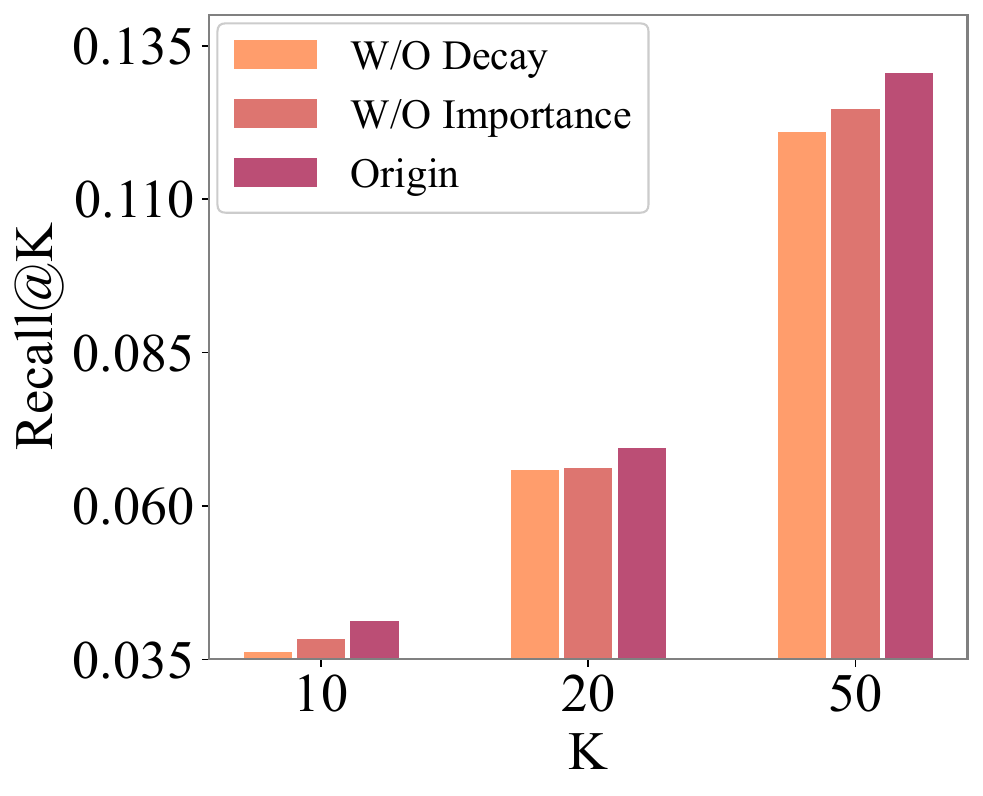}
    \caption{Recall on Yelp2018}
\end{subfigure}
\begin{subfigure}{0.32\linewidth}
    \includegraphics[width=\linewidth]{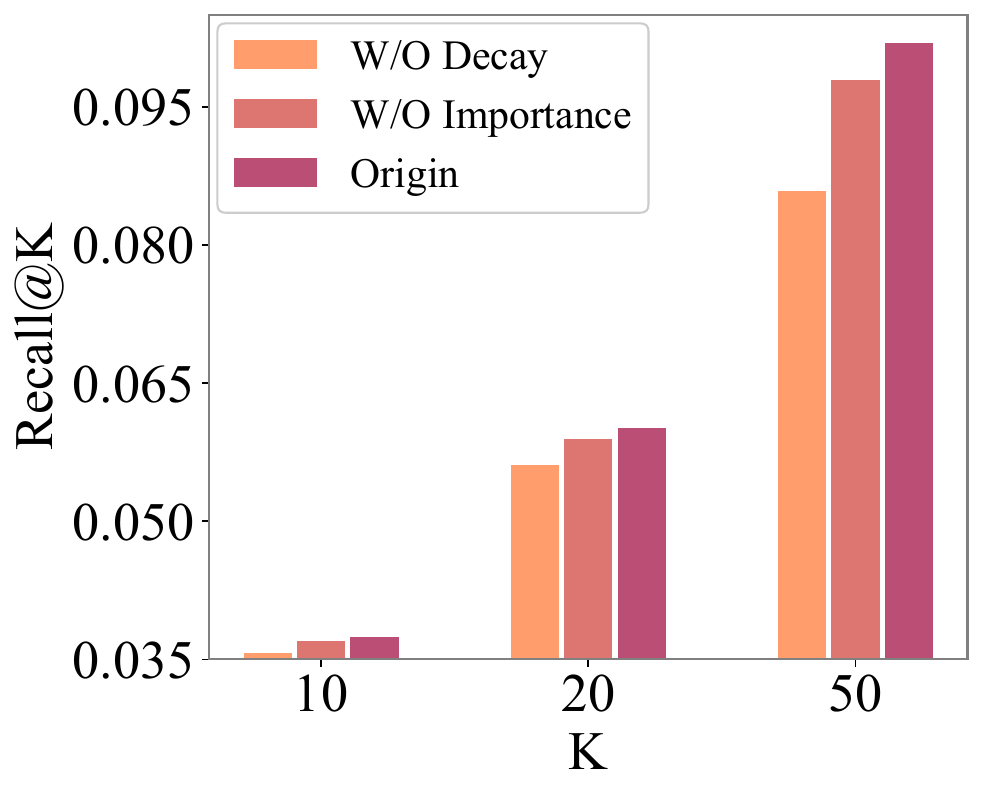}
    \caption{Recall on Amazon-Book}
\end{subfigure}

\begin{subfigure}{0.32\linewidth}
    \includegraphics[width=\linewidth]{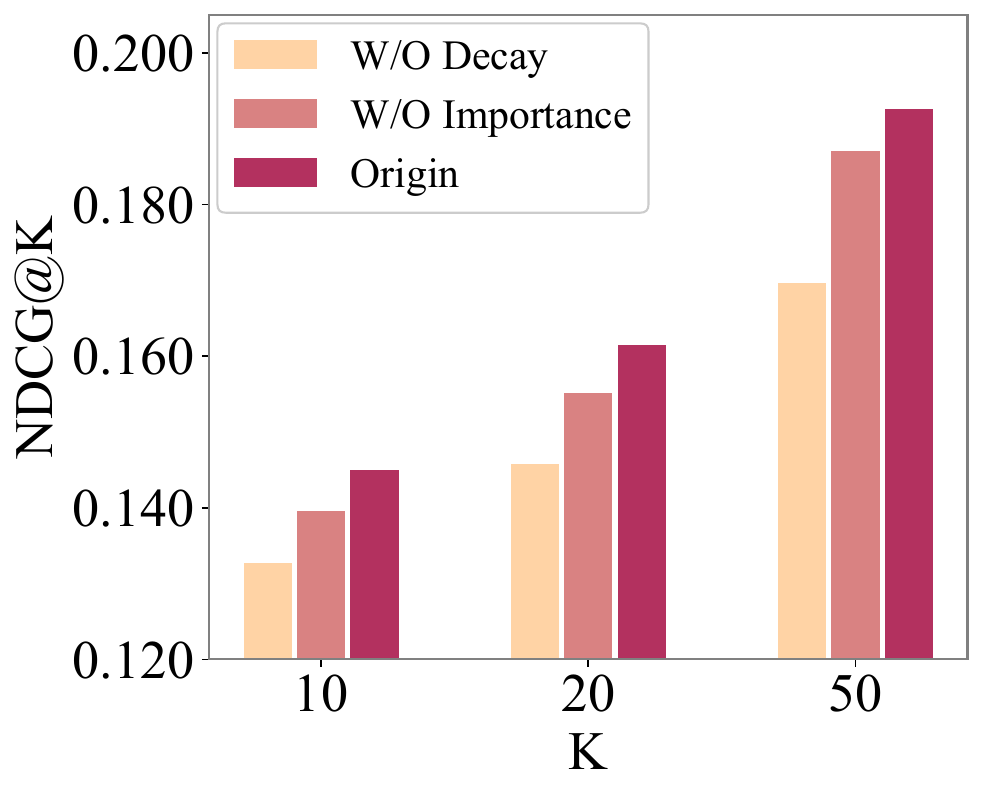}
    \caption{NDCG on Gowalla}
\end{subfigure}
\begin{subfigure}{0.32\linewidth}
    \includegraphics[width=\linewidth]{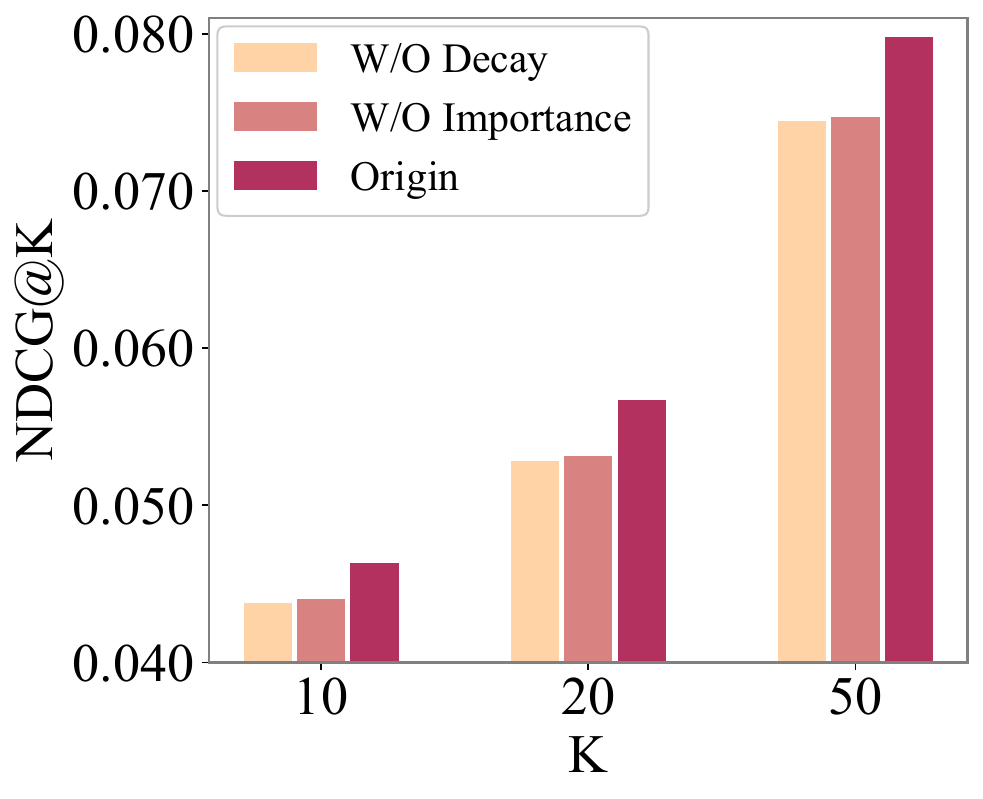}
    \caption{NDCG on Yelp2018}
\end{subfigure}
\begin{subfigure}{0.32\linewidth}
    \includegraphics[width=\linewidth]{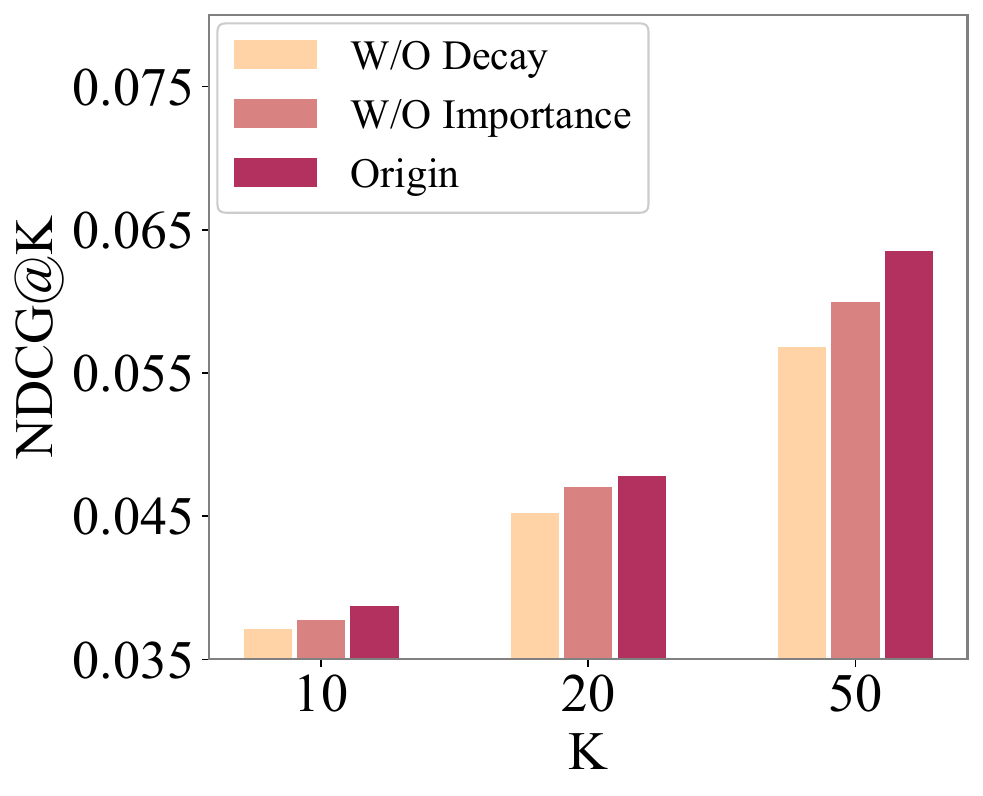}
    \caption{NDCG on Amazon-Book}
\end{subfigure}

\caption{Ablation performance w.r.t. personalized decay and importance sampling.} 
\label{fig:ablation}
\end{figure}
The forward interests burn-down process of \method{} relies on two modules to encourage the personality in model optimization: a graph convolution kernel and an importance sampling weight. To validate their effectiveness, we conduct ablation studies on these two components and present the model performance in Figure \ref{fig:ablation}. The results reveal the following insights:
\begin{itemize}[leftmargin=*]
    \item Both the decay factor and the importance weight are crucial for \method{} to achieve promising performance. These two key components in the burn-down process , contributing to enhanced performance. The decay factor introduces personalized item-wise information into the diffusion process, while the importance weight imparts time-wise significance, thereby improving model optimization.
    \item  Specifically, the personalized decay factor plays a more vital role in the optimization phase of \method{}. This highlights that models adept at handling personalized item preferences are likely to achieve superior performance, aligning with the intuitive understanding that collaborative relationships between items are fundamental in recommendation systems.
\end{itemize}
\input{Tables/ablation_decay}
\R{
\subsubsection{Influence of Decay Schemes}

The proposed \method{} models interest evolution using a stochastic decay path based on binomial sampling with an exponential form. However, similar effects may be achieved using alternative decay designs. To assess the benefit of the adopted random binomial decay, we evaluate \method{} under several deterministic and stochastic decay schemes during training:
\begin{itemize}[leftmargin=*]
    \item \textbf{Exponential decay} applies a deterministic decay, where the interest representation at step $t$ is obtained with $X_u^{(t)}=X_u^{(0)}\cdot e^{-\lambda t}$, where $\lambda$ controls the decay rate.
    \item \textbf{Power decay} replaces the exponential form with a power-law probability, using a binomial process $X_u^{(t)}\sim Binomial(X_u^{(0)},(1+t)^{-\alpha})$, where $\alpha>0$ is a constant.
    \item \textbf{Linear decay} follows a similar construction but uses a linear decay rate, given by $\max\{0,1-\beta t\}$.
    \item \textbf{Gaussian rescaled decay} adopts a diffusion-inspired approach, in which the interest vector is scaled as $X_u^{(t)}=s_t\cdot X_u^{(0)}$. The scaling factor $s_t\sim TN_{[0,1]^m}(\alpha_t,\sigma^2_t)$ is sampled from a truncated Gaussian distribution over, with a linear noise schedule following DDPM\cite{ho2020denoising}.
\end{itemize}

Based on the experimental results in Table \ref{tab:abla_decay}, we can conclude that multiple decay schemes can be used to build diffusion-based recommender models. However, the proposed interest decay process consistently outperforms the alternatives, highlighting the benefit of a discrete and personalized design. Specifically, stochastic decay schemes achieve better performance than deterministic ones, indicating that incorporating randomness into the diffusion process leads to more effective learning.
}

\subsection{Hyperparameter Analysis}

The proposed \method{} involves several predefined hyperparameters that guide the training and inference procedures, significantly impacting recommendation performance. To reveal the sensitivity of \method{} to these parameters, we conduct a series of parameter analysis on \textbf{Gowalla} and \textbf{Yelp2018} datasets.

\subsubsection{Influence of Sampling Steps}

\begin{figure}
\centering
\begin{subfigure}{0.32\linewidth}
    \includegraphics[width=\linewidth]{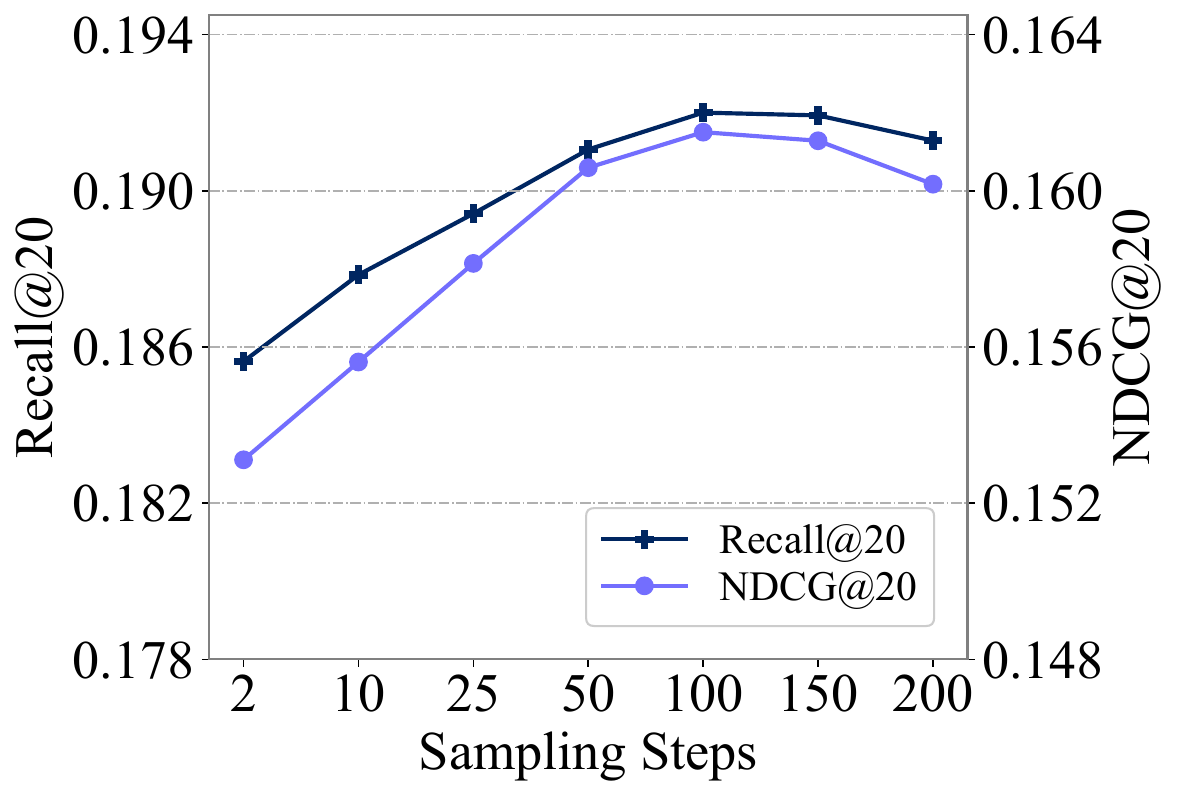}
    \caption{Gowalla}
    \label{fig:sample_stap:a}
\end{subfigure}
\begin{subfigure}{0.32\linewidth}
    \includegraphics[width=\linewidth]{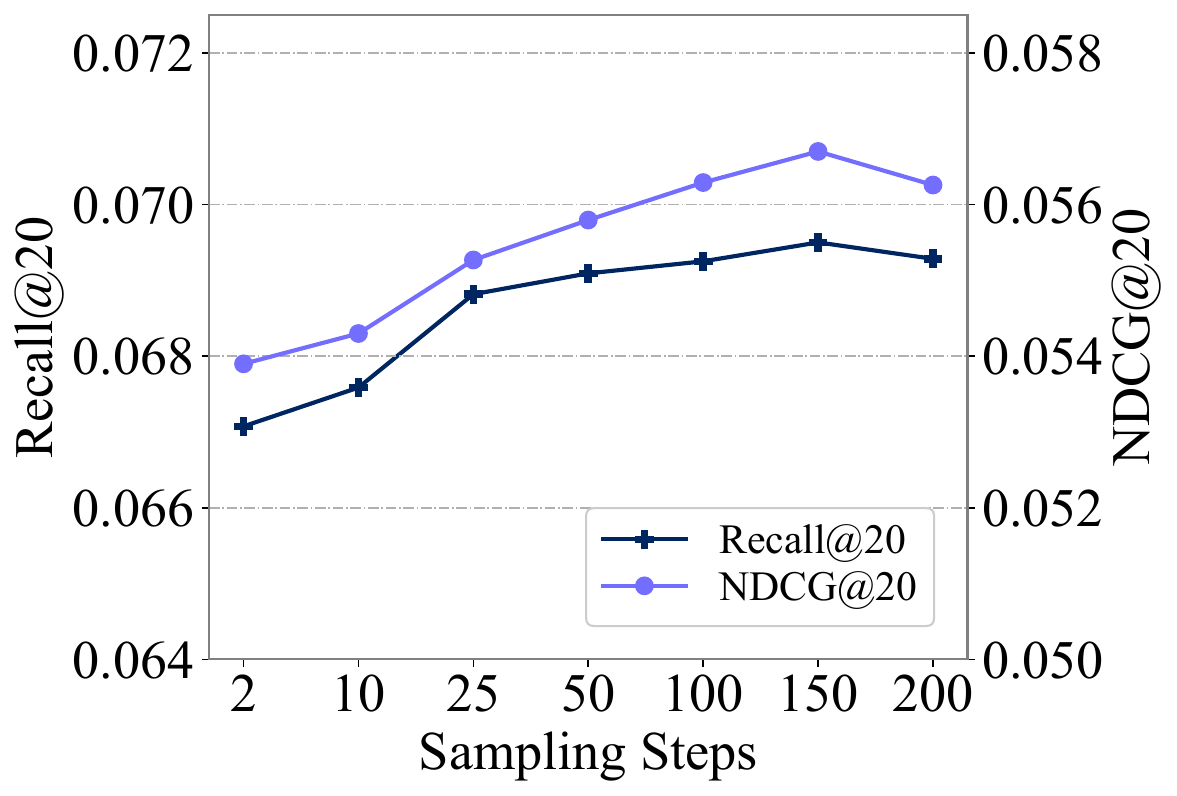}
    \caption{Yelp2018}
    \label{fig:sample_stap:b}
\end{subfigure}
\begin{subfigure}{0.32\linewidth}
    \includegraphics[width=\linewidth]{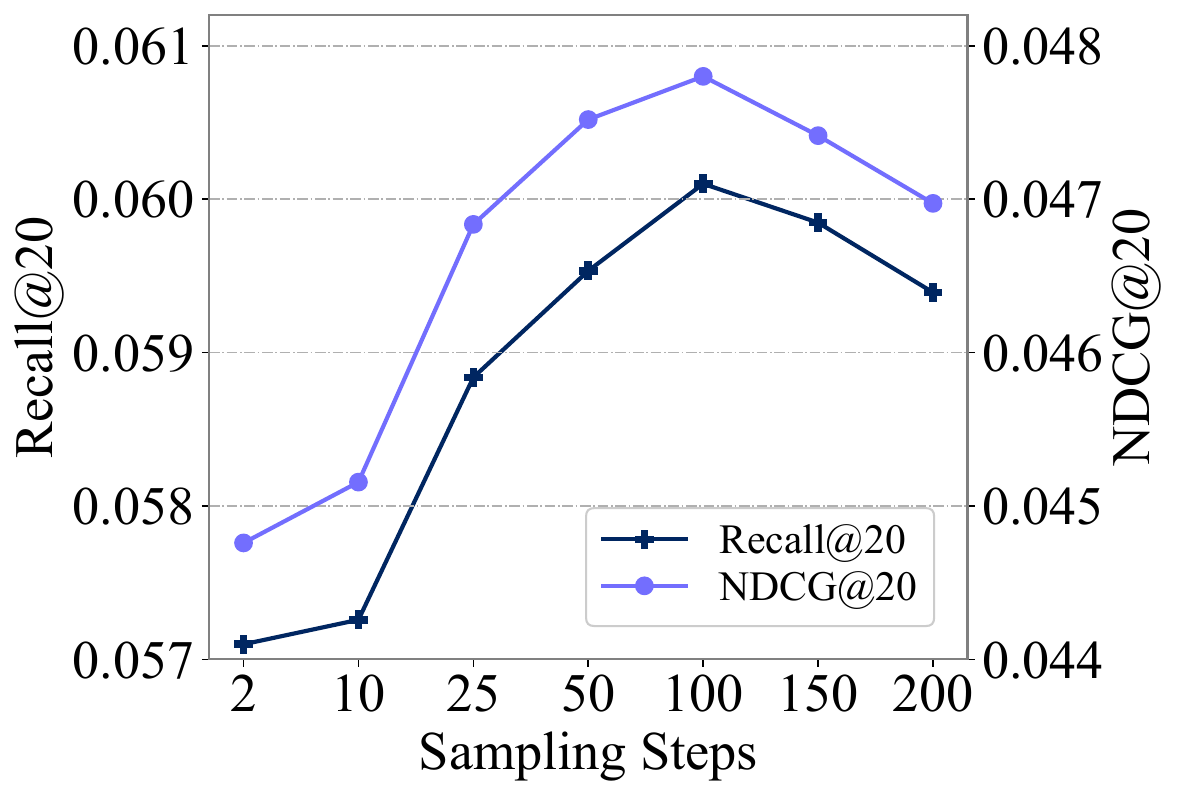}
    \caption{Yelp2018}
    \label{fig:sample_stap:c}
\end{subfigure}

\caption{Model performance w.r.t. diffusion sampling steps.} 
\label{fig:sample_step}
\end{figure}
The diffusion sampling time steps $T/\Delta t$ defines the granularity of the diffusion process. To assess the effect of the sampling granularity, we conduct two sets of parameter studies. We tune $\Delta t$ to investigate the influence of diffusion time steps and the results are reported in Figure \ref{fig:sample_step}. The outcomes reveal that with increasing sampling steps, the performance of \method{} benefits from informative diffused samples within a finely-grained time-span, leading to a corresponding improvement. However, the recommendation performance stops growing as the granularity is too fine, reaching a point where the model isn't able to learn more information from the process.



\subsubsection{Impact of Stage Number}

\begin{figure}
\centering
\begin{subfigure}{0.32\linewidth}
    \includegraphics[width=\linewidth]{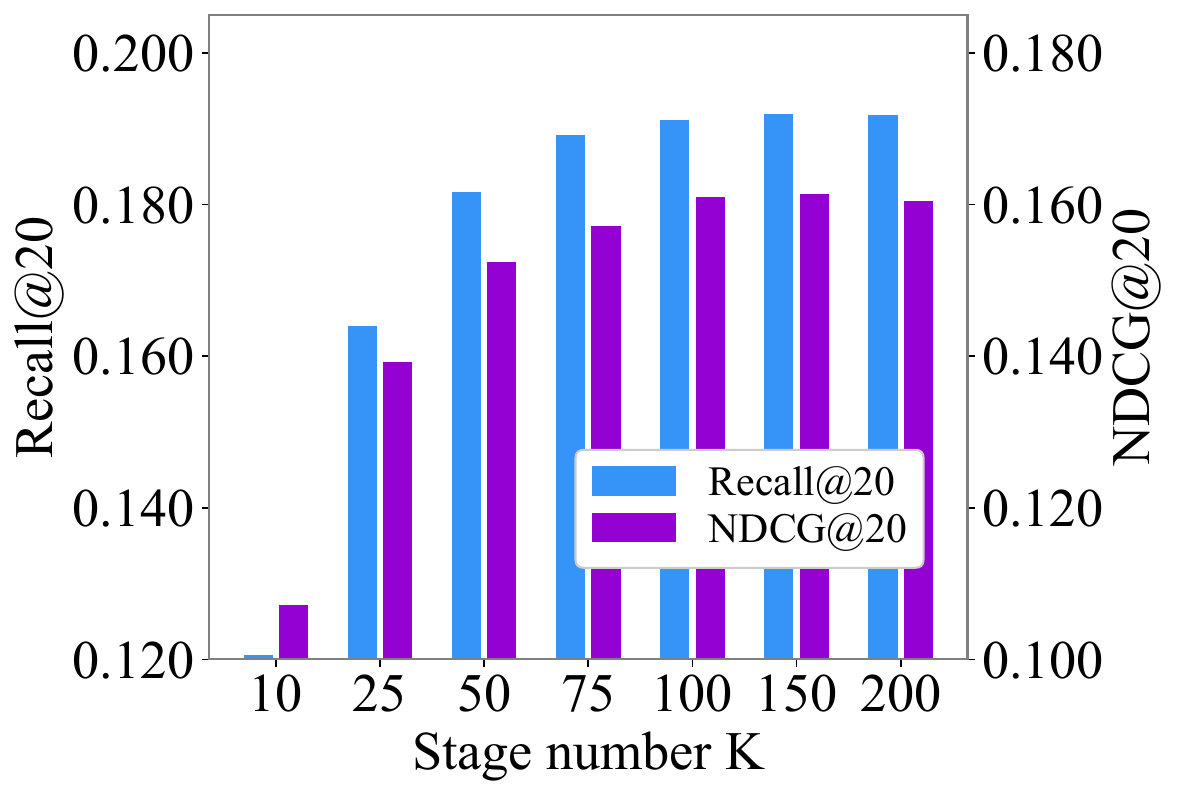}
    \caption{Gowalla}
\end{subfigure}
\begin{subfigure}{0.32\linewidth}
    \includegraphics[width=\linewidth]{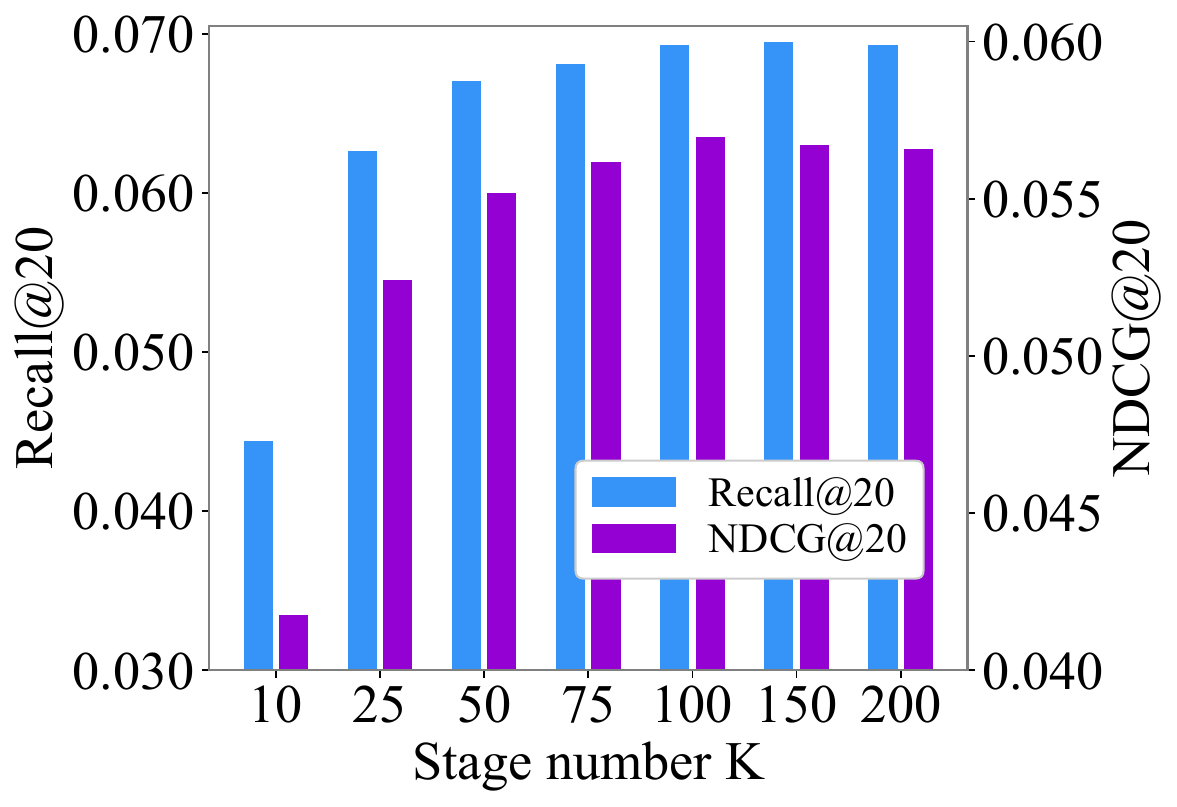}
    \caption{Yelp2018}
\end{subfigure}
\begin{subfigure}{0.32\linewidth}
    \includegraphics[width=\linewidth]{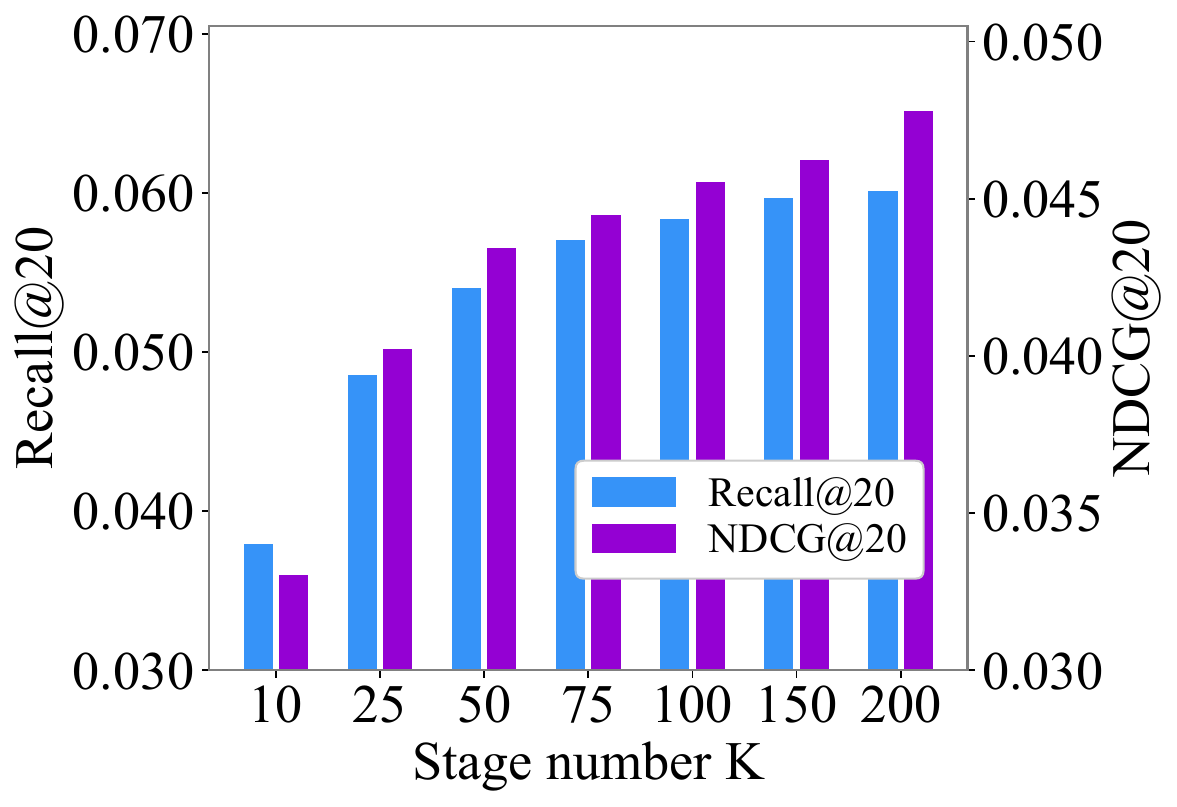}
    \caption{Amazon-Book}
\end{subfigure}

\caption{Model performance w.r.t. granularity number.} 
\label{fig:stage_num}
\end{figure}
The proposed interests burn-down process relies on a multi-stage interests vectors as diffused samples, making the predefined stage number $K$ a crucial factor influencing model performance. The model performance concerning the maximum stage number $K$ is illustrated in Figure \ref{fig:stage_num}. The results highlight the significance of employing multiple stage-like data samples in \method{}. A more finely grained stages staging approach proves beneficial for the model to effectively capture the dynamic decay of user interests. However, it is noteworthy that the optimal stage number varies across different datasets, which indicates that the granularity of interests is highly relevant to practical scenarios.

\subsubsection{Effect of Personalized Decay Weight}

\begin{figure}
\centering
\begin{subfigure}{0.31\linewidth}
    \includegraphics[width=\linewidth]{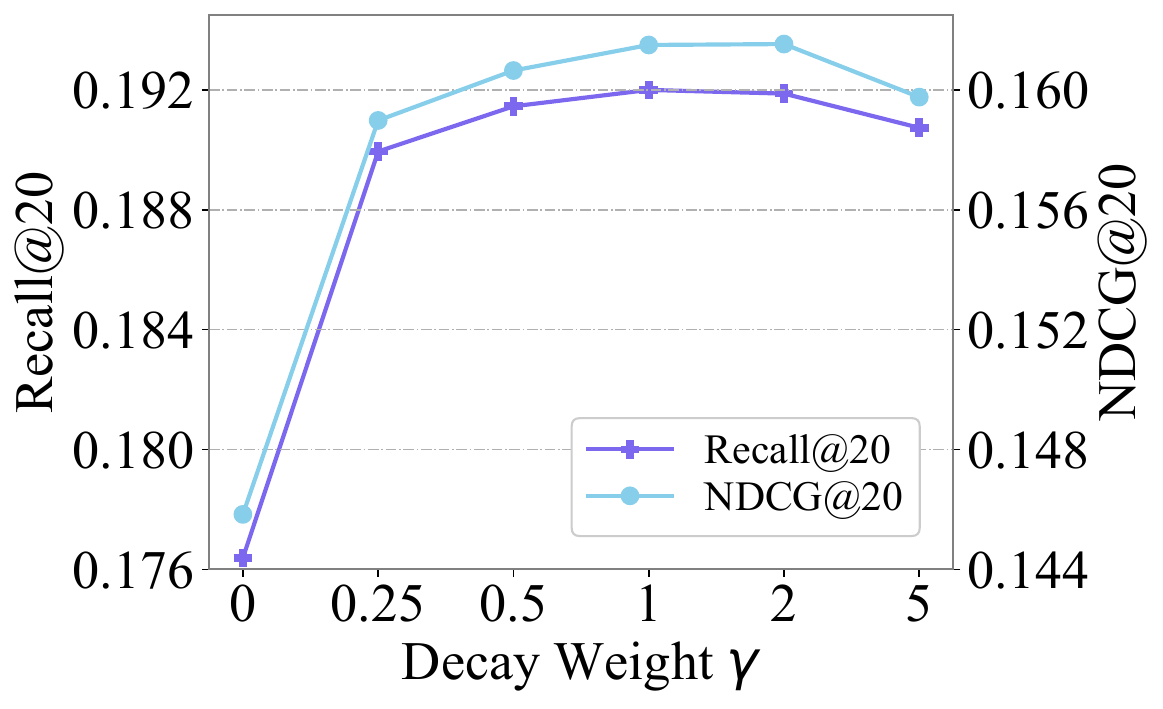}
    \caption{Gowalla}
\end{subfigure}
\begin{subfigure}{0.31\linewidth}
    \includegraphics[width=\linewidth]{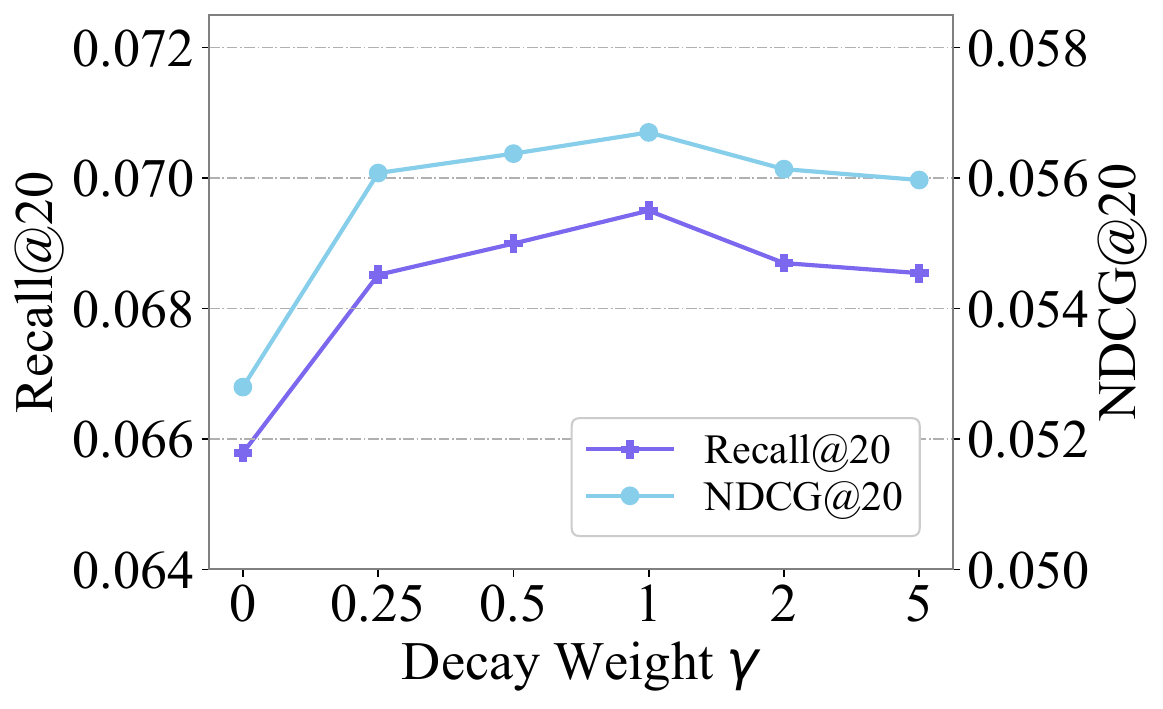}
    \caption{Yelp2018}
\end{subfigure}
\begin{subfigure}{0.31\linewidth}
    \includegraphics[width=\linewidth]{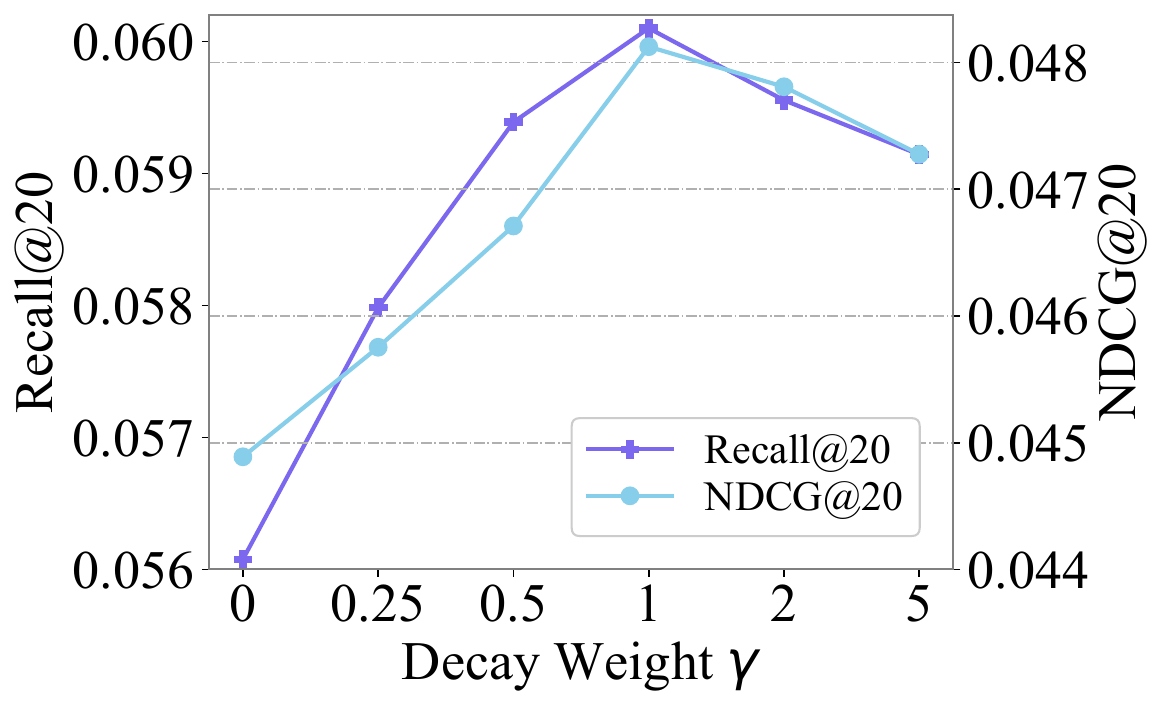}
    \caption{Amazon-Book}
\end{subfigure}

\caption{Model performance w.r.t. personalized decay.} 
\label{fig:decay_weight}
\end{figure}
The influence of personalized decay factor is steered by a weight parameter $\gamma$, which controls the collaborative effect ratio in the interests burn-down process. To investigate the influence of the decay factor, we vary its weight from none ($\gamma=0$) to heavy ($\gamma=5$) in a parameter study. The results in Figure \ref{fig:decay_weight} reveal that the weight of decay factor within specific ranges enhances the model performance. On the other hand, an extreme decay weight would lead to sub-optimal results due to the lack of randomness in the diffusion process, preventing the model from learning rich information from the burnt-down interests.

\subsubsection{Influence of Sampling Time}

\begin{figure}
\centering
\begin{subfigure}{0.31\linewidth}
    \includegraphics[width=\linewidth]{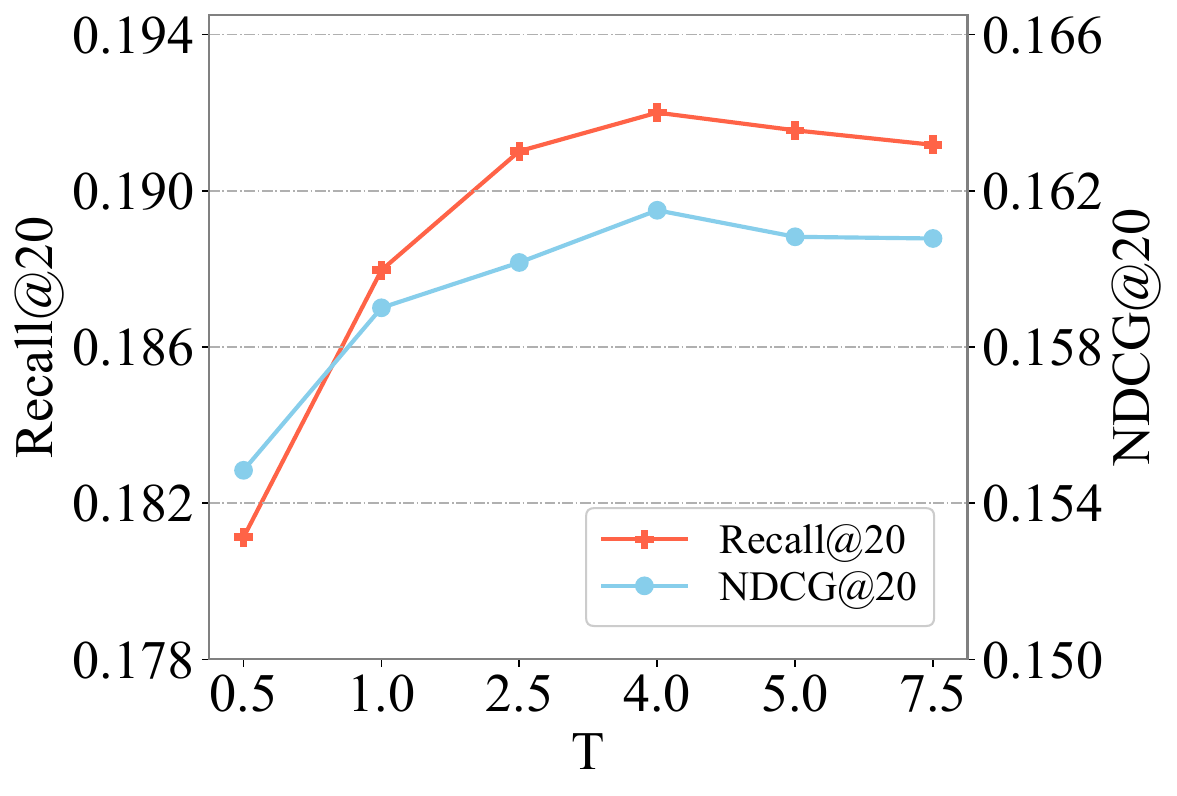}
\end{subfigure}
\begin{subfigure}{0.31\linewidth}
    \includegraphics[width=\linewidth]{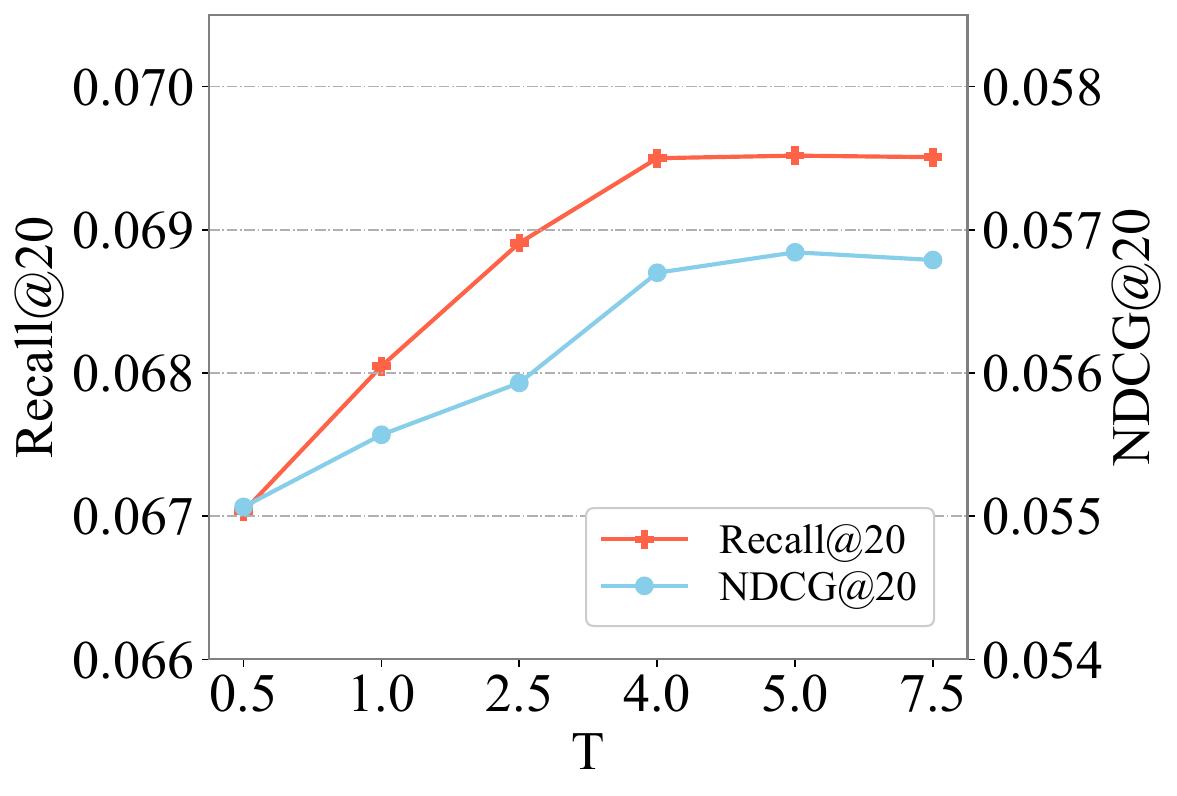}
\end{subfigure}
\begin{subfigure}{0.31\linewidth}
    \includegraphics[width=\linewidth]{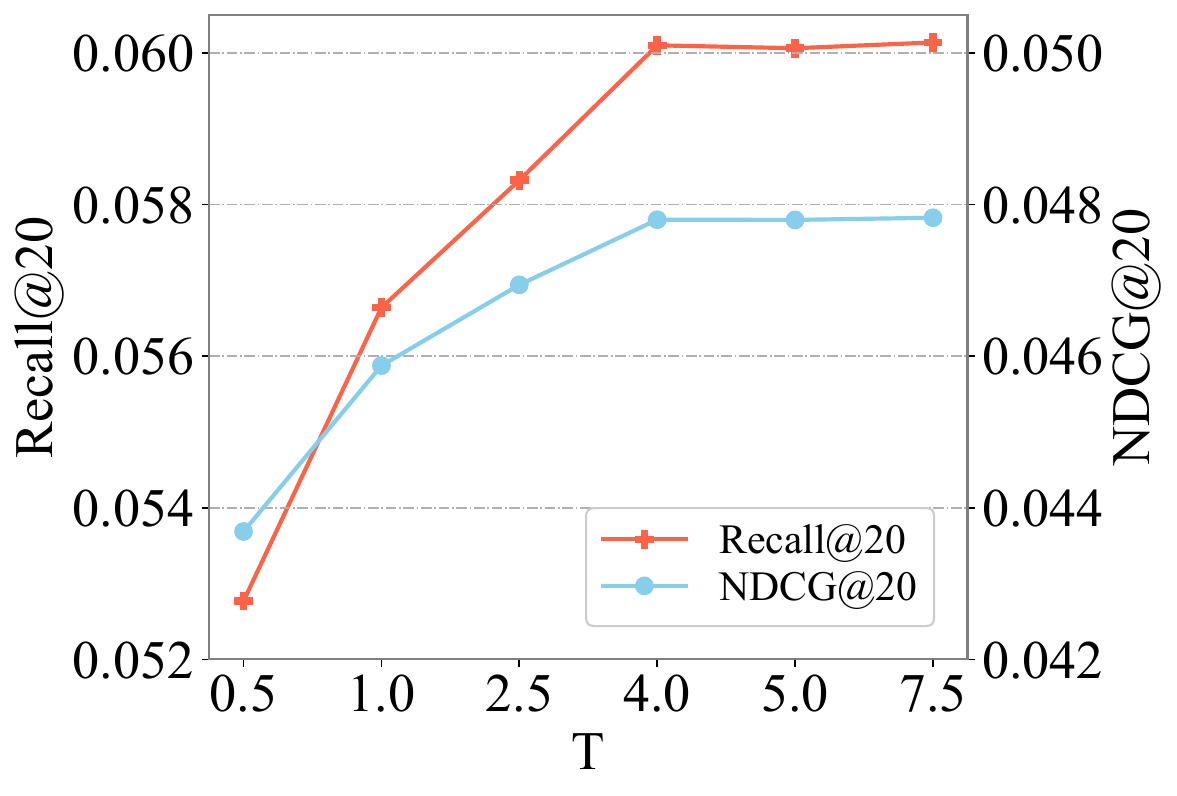}
\end{subfigure}
\caption{Model performance w.r.t. sampling time.} 
\label{fig:sample_time}
\end{figure}

The interests burn-down diffusion process in \method{} is scaled by the diffusion time parameter $T$ and $T'$. By default, we choose a moderate $T=T'$ so that all the interests have almost decayed to zero when the diffusion reaches $X_u^{(T)}$. To further investigate the effect of different sampling time, we record model performance with respect to $T$ in Figure \ref{fig:sample_time}. It can be observed that a larger sampling time gap $T$ would bring benefits for providing more diffused samples of interests. However, an extreme $T$ would cause the performance decline since most of the interests have been decayed to zero.

\subsection{In-depth Study}


\subsubsection{Training Curves}

\begin{figure}
\centering
Training curves on \textbf{Gowalla}.\ \ \ \ \ \ \ \ \ \ \ \ \ \ \ \ \ \ \ \ \ \ \ \ \ Training curves on \textbf{Yelp2018}.

\begin{subfigure}{0.24\linewidth}
    \includegraphics[width=\linewidth]{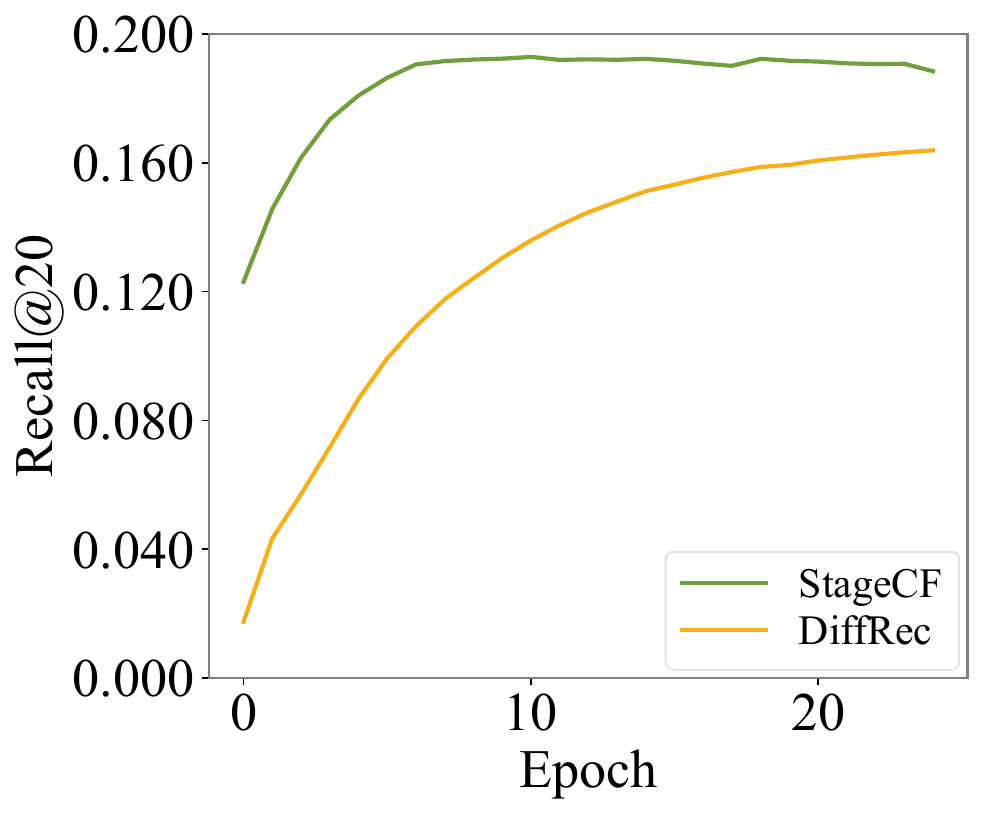}
\end{subfigure}
\begin{subfigure}{0.24\linewidth}
    \includegraphics[width=\linewidth]{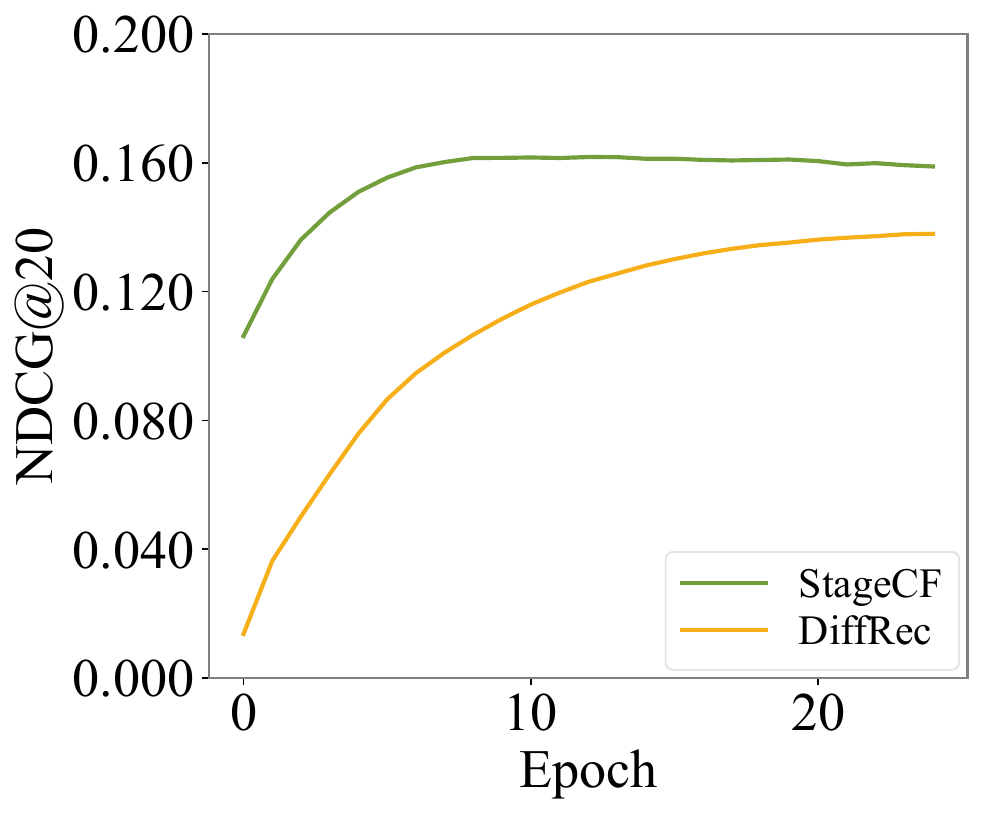}
\end{subfigure}
\begin{subfigure}{0.24\linewidth}
    \includegraphics[width=\linewidth]{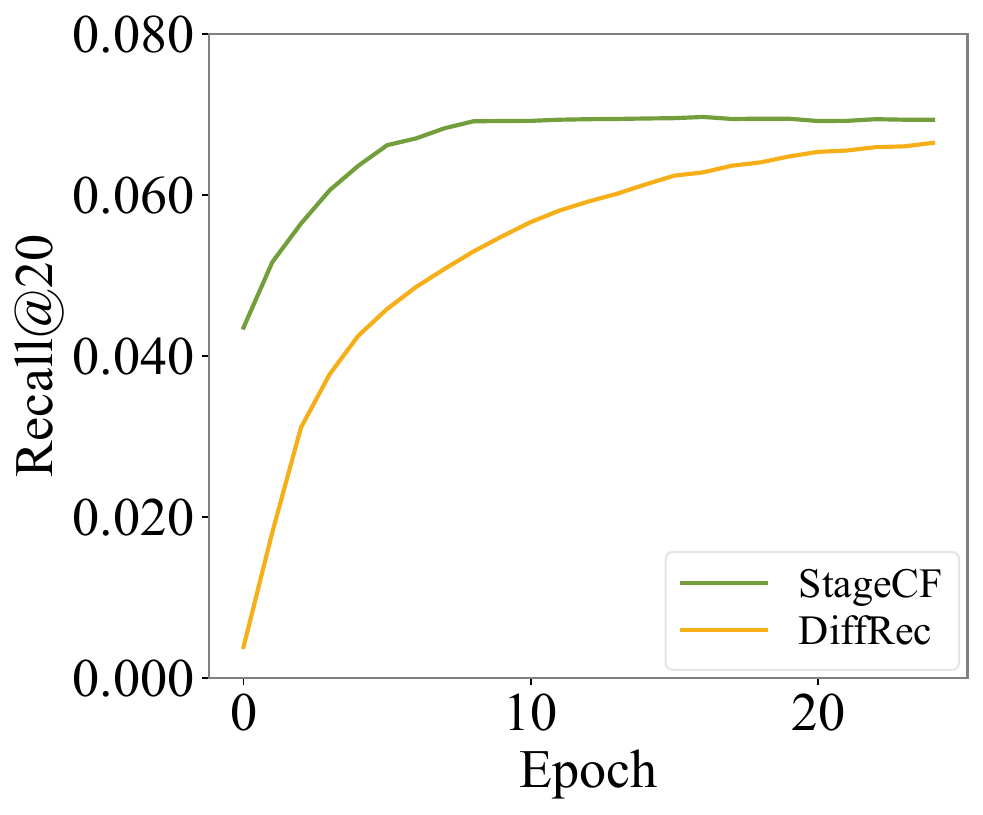}
\end{subfigure}
\begin{subfigure}{0.24\linewidth}
    \includegraphics[width=\linewidth]{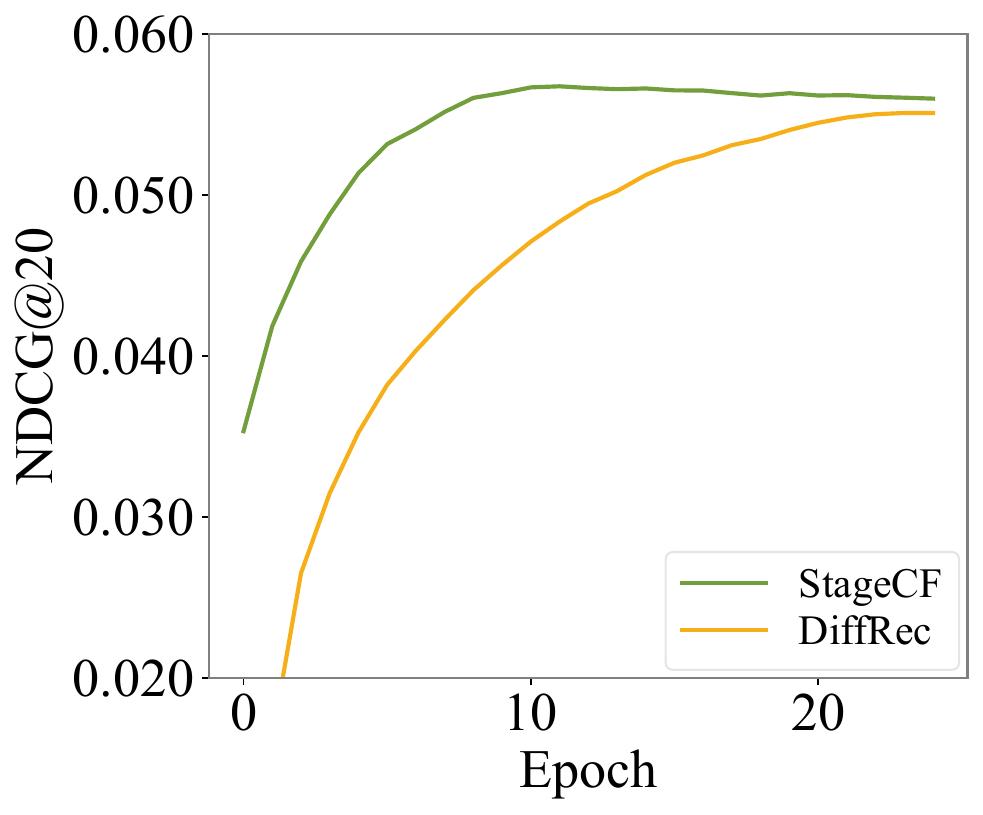}
\end{subfigure}

\caption{Training curves of \method{} and DiffRec.} 
\label{fig:curve}
\end{figure}
To intuitively illustrate the model performance of \method{} and relative diffusion-based methods during training, we visualize the training curve of \method{} and DiffRec in Figure \ref{fig:curve}. Under the same setting of learning rate $lr=5\times10^{-5}$, from the training curves we can observe that the interests burn-down process helps \method{} converge faster than Gaussian diffusion process, reflecting the informative information brought by personalized interests decay. Specifically, \method{} achieves faster convergence, as well as higher performance around each epoch of training, showing the superiority of the interests burn-down process.

\subsubsection{Training and Inference Efficiency}
\input{Tables/efficiency}
\R{While the execution efficiency is an important factor in assessing the practical value of recommender systems in real-world settings, we evaluate both the training and inference efficiency of \method{} and compare it with several baseline methods. Specifically, Table \ref{tab:effiecency} reports the training time required to reach convergence and the inference time over the entire test set. The results show that \method{} has slower inference than methods with fewer forward passes, such as LightGCN, MacridVAE, and FlowCF. However, its inference efficiency is comparable to that of other diffusion-based collaborative filtering models, including DiffRec and DDRM. Notably, among sampling-based methods with similar inference costs, \method{} converges substantially faster during training. Overall, these findings indicate that the proposed discrete interest decay process leads to a more effective and efficient training objective, while preserving competitive inference efficiency.
}

\subsubsection{Group-wise Recommendation Performance}

\begin{figure}
\centering
\begin{subfigure}{0.32\linewidth}
    \includegraphics[width=\linewidth]{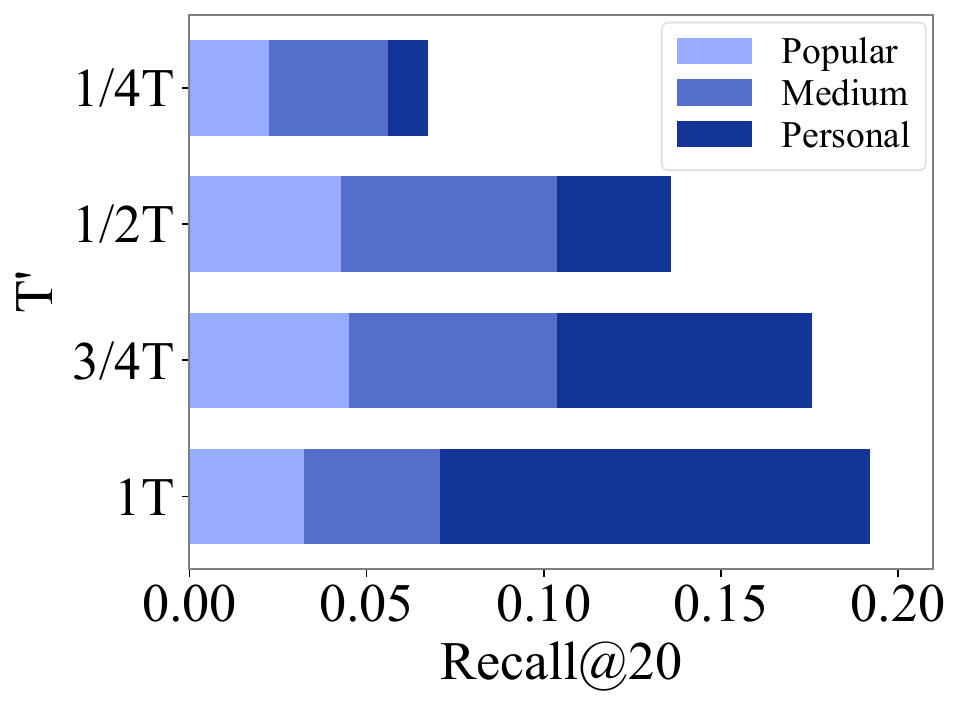}
    \caption{Gowalla}
\end{subfigure}
\begin{subfigure}{0.32\linewidth}
    \includegraphics[width=\linewidth]{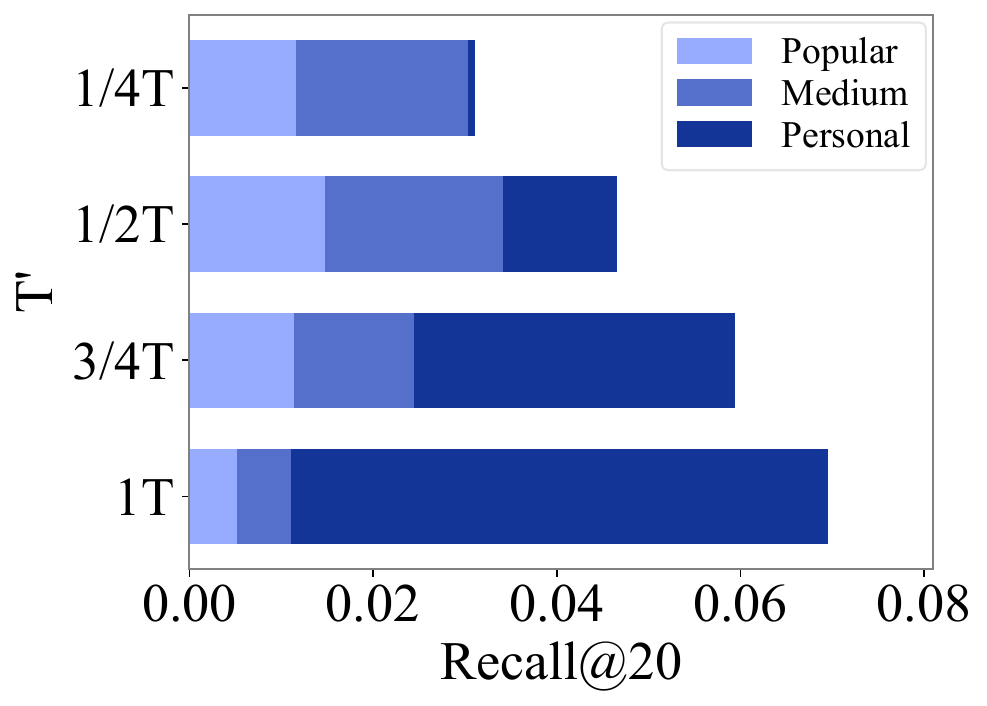}
    \caption{Yelp2018}
\end{subfigure}
\begin{subfigure}{0.32\linewidth}
    \includegraphics[width=\linewidth]{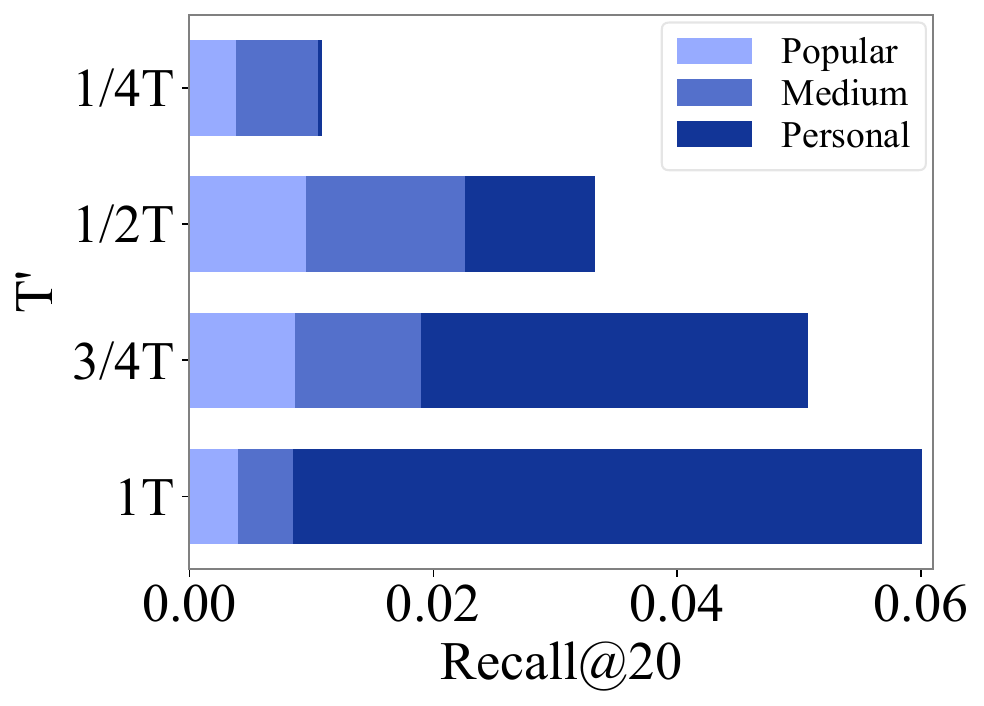}
    \caption{Amazon-Book}
\end{subfigure}

\caption{Model performance on different user groups w.r.t. sampling step.} 
\label{fig:popularity}
\end{figure}
Recall that the recommendation procedure of \method{} involves two fundamental factors, which are the burnt-up interests updated during the sampling process, and the personalized burn-up transition rate provided by the score estimation network.
To assess whether the personalized interactions have been correctly reconstructed through the reverse burn-up process, we conduct in-depth study to intuitively illustrate the personalized interests during the sampling process. To achieve this, we evaluate the recommendation results at different stages of the sampling process to test model performance on separate item groups. 

Specifically, we categorize all the items into three groups according to their popularity: popular items (items with top 33\% interactions), medium items, and personal items (bottom 33\% interactions). We then evaluate the model performance on each item group respectively. From results in Figure \ref{fig:popularity} we can observe that, from the least personal interactions (interaction with popular items) to the most personal interactions (interaction with personal items), as the sampling process progresses from interactions with popular items to interactions with personal items, the recommendation results tend to increasingly reflect personalized choices. This observation suggests that the diffusion sampling process contributes to generating recommendations with more personalized outcomes.

\subsubsection{Influence of Training and Sampling Approach}
\input{Tables/indepth}

While the training and sampling processes of \method{} have been defined in previous sections, recent research on black-out diffusion \cite{santos2023blackout} has theoretically analyzed possible optimization and sampling options. Specifically, the transition rate of the diffusion process can be prescribed as instantaneous time-dependent or defined on finite-time leap, resulting in different forms of optimization target. Specifically, the finite-time form of learning objective is formulated as:
\begin{equation}
\label{eq:loss_fn_finite}
    \mathcal{L}_{finite}=t\cdot e^{-t}(q_u(t)-(X_u^{(0)}-X_u^{(t)})\log q_u(t)).
\end{equation}
which differs from the default objective in Equation \ref{eq:loss_fn} by adding an extra weight on time step $t$. We term this method as \textit{Finite-time} form of the optimization target, and the original one in Equation \ref{eq:loss_fn} is termed as \textit{Instantaneous} form.

Similarly, there is another substitute of the sampling process in Equation \ref{eq:backward_distribution} by constructing sample functions from Poisson distributions, formulated as:
\begin{equation}
\label{eq:backward_distribution_poisson}
    X_{u}^{(t-\Delta t)}-X_u^{(t)}\sim Poisson(\frac{e^{-t}}{1-e^{-t}}s_u(t)),
\end{equation}
which performs the sampling strategy based on the \textit{Poisson} random numbers, and the original sampling process in Equation \ref{eq:backward_distribution} is identified as \textit{Binomial-bridge} form.

These possible options of optimization and sampling methods make up to four total variants of training and recommendation scheme and can also fit to the current \method{} framework. To investigate \method{}'s behavior under different assumptions of time intervals and approximations, we conduct in-depth study and the model performance is reported in Table \ref{tab:indepth}. From the experimental results we can observe that:

\begin{itemize}[leftmargin=*]
    \item The default setting adopted by \method{}, i.e. the last line in Table \ref{tab:indepth} shows the optimal performance, compared with other optimization and recommendation schemes.
    \item In general, the finite-time form of the objective function achieves similar recommendation performance as the instantaneous form of training objective.
\end{itemize}


%% file: Tables/statistics.tex
\begin{table}
\centering
\caption{Descriptive statistics of the used datasets, where "\#Avg. per User" denotes average observed interaction number of each user.}
\label{tab:statics}
\setlength{\tabcolsep}{3pt}
\begin{tabular}{ccccc} 
\toprule 
\textbf{Dataset} & \#User & \#Item & \#Interactions & \#Avg. per User\\
\midrule
Gowalla & 29,858 & 40,981 & 1,027,370 & 34.41 \\
Yelp2018 & 31,668 & 38,048 & 1,561,406 & 49.31 \\
Amazon-Book & 52,643 & 91,599 & 2,984,108 & 56.68 \\
\bottomrule 
\end{tabular}
\end{table}

%% file: Tables/main_result.tex
\begin{table*}
\centering
\caption{The test results of \method{} and all baselines, where R@K and N@K are short for Recall@K and NDCG@K. The highest performance is emphasized with bold font and the second highest is marked with underlines. Results marked with $^\star$ indicate statistical significance based on a paired t-test with $p < 0.01$.}
\label{tab:overall}
\renewcommand{\arraystretch}{1.1}
\resizebox{\linewidth}{!}{
\begin{tabular}{cc|ccc|ccc|cccc|cc}
\toprule 
Dataset & Metrics & \R{SLIM} & \RR{iALS} & LightGCN & MultDAE & MultVAE & MacridVAE & DiffRec & \R{FlowCF} & DDRM  & \method{} & Improv. \\

\hline
\multirow{6}{*}{\shortstack{\textbf{Gowalla}}}
& R@10 & 0.1095 & 0.1129 & {0.1288} & 0.1055 & 0.1025 & 0.1012 & 0.1144 & 0.1160 & \underline{0.1288} & \textbf{0.1351}$^\star$ & 4.89\% \\
& R@20 & 0.1560 & 0.1619 & {0.1831} & 0.1518 & 0.1481 & 0.1490 & 0.1659 & 0.1694 & \underline{0.1829} & \textbf{0.1915}$^\star$ & 4.70\% \\
& R@50 & 0.2380 & 0.2515 & {0.2689} & 0.2406 & 0.2344 & 0.2485 & 0.2561 & 0.2680 & \underline{0.2720} & \textbf{0.2878}$^\star$ & 5.81\% \\
& N@10 & 0.1169 & 0.1223 & {0.1399} & 0.1109 & 0.1100 & 0.0912 & 0.1253 & 0.1269 & \underline{0.1399} & \textbf{0.1455}$^\star$ & 4.00\% \\
& N@20 & 0.1303 & 0.1369 & {0.1551} & 0.1233 & 0.1229 & 0.1083 & 0.1382 & 0.1411 & \underline{0.1546} & \textbf{0.1613}$^\star$ & 4.33\% \\
& N@50 &  0.1558 & 0.1652 & {0.1842} & 0.1519 & 0.1461 & 0.1392 & 0.1671 & 0.1718 & \underline{0.1851} & \textbf{0.1910}$^\star$ & 3.19\% \\

\hline
\multirow{6}{*}{\shortstack{\textbf{Yelp2018}}}
& R@10  & 0.0369 & 0.0385 & 0.0376 & 0.0365 & 0.0371 & 0.0365 & {0.0382} & \underline{0.0390} & {0.0388} & \textbf{0.0414}$^\star$ & 6.15\% \\
& R@20 & 0.0610 & 0.0649 & 0.0642 & 0.0641 & 0.0633 & 0.0641 & {0.0660} & \underline{0.0659} & 0.0652  & \textbf{0.0696}$^\star$ & 5.51\% \\
& R@50 & 0.1123 & 0.1211 & 0.1231 & 0.1244 & 0.1232 & 0.1244 & {0.1231} & \underline{0.1246} & {0.1243} & \textbf{0.1304}$^\star$ & 4.65\% \\
& N@10 & 0.0438 & 0.0446 & 0.0430 & 0.0421 & 0.0426 & 0.0405 & {0.0459} & \underline{0.0452} & 0.0440 & \textbf{0.0466}$^\star$ & 3.10\% \\
& N@20 & 0.0521 & 0.0508 & 0.0529 & 0.0519 & 0.0508 & 0.0512 & {0.0545}& \underline{0.0546} & 0.0542 & \textbf{0.0570}$^\star$ & 4.40\% \\
& N@50 & 0.0710 & 0.0722 & 0.0748 & 0.0755 & 0.0731 & 0.0733 & {0.0742} & \underline{0.0759} & 0.0755 & \textbf{0.0800}$^\star$ & 5.40\% \\

\hline
\multirow{6}{*}{\shortstack{\textbf{Amazon-Book}}}
& R@10 & 0.0310 & 0.0282 & 0.0242 & 0.0282 & 0.0265 & 0.0262 & {0.0319} & \underline{0.0331} & 0.0267 & \textbf{0.0380}$^\star$ & 14.80\% \\
& R@20 & 0.0506 &  0.0477 & 0.0417 & 0.0465 & 0.0448 & 0.0441 & {0.0520} & \underline{0.0529} & 0.0432 & \textbf{0.0604}$^\star$ & 14.18\% \\
& R@50 & 0.0869 & 0.0885 & 0.0814 & 0.0850 & 0.0811 & 0.0822 & {0.0902} & \underline{0.0907} & 0.0806 & \textbf{0.1033}$^\star$ & 13.89\% \\
& N@10 & 0.0304 & 0.0297 & 0.0251 & 0.0301 & 0.0291 & 0.0281 & {0.0332} & \underline{0.0344} & 0.0272 & \textbf{0.0390}$^\star$ & 13.37\% \\
& N@20 & 0.0406 & 0.0374 & 0.0322 & 0.0382 & 0.0364 & 0.0352 & {0.0419} & \underline{0.0435} & 0.0336  & \textbf{0.0484}$^\star$ & 11.26\% \\
& N@50 & 0.0571 & 0.0534 & 0.0469 & 0.0535 & 0.0508 & 0.0499 & {0.0574} & \underline{0.0590} & 0.0474  & \textbf{0.0639}$^\star$ & 8.31\% \\

\bottomrule 
\end{tabular}
}

\end{table*}

%% file: Tables/ablation_diffusion.tex
\begin{table*}
\centering
\caption{The ablation results of different diffusion scheme, as well as the influence of different styles on combining diffusion process into recommendation models. The "R@K" and "N@K" represents Recall and NDCG respectively.} 
\label{tab:abla_diff} 

\resizebox{\linewidth}{!}{
\begin{tabular}{cc cccccccccccc}
\toprule 
\multirow{2}{*}{Method} & \multirow{2}{*}{Diffusion Scheme} & \multicolumn{4}{c}{\textbf{Gowalla}} & \multicolumn{4}{c}{\textbf{Yelp2018}} & \multicolumn{4}{c}{\textbf{Amazon-Book}} \\ 
\cmidrule[0.5pt](lr){3-6}\cmidrule[0.5pt](lr){7-10}\cmidrule[0.5pt](lr){11-14}
& & R@10 & R@20 & N@10 & N@20 & R@10 & R@20 & N@10 & N@20 & R@10 & R@20 & N@10 & N@20  \\
\midrule 
\textbf{M0} & N/A & 0.1107 & 0.1636 & 0.1225 & 0.1321 & 0.0360 & 0.0636 & 0.0435 & 0.0541 & 0.0306 & 0.0481 & 0.0319 & 0.0394 \\
\midrule

\textbf{\RR{M1}} & Burn-down & 0.0542 & 0.1052 & 0.0463 & 0.0805 & 0.0157 & 0.0434 & 0.0202 & 0.0395 & 0.0124 & 0.0276 & 0.0106 & 0.0249 \\
\midrule
\multirow{2}{*}{\textbf{M2}}
& Gaussian & 0.1160 & 0.1652 & 0.1242 & 0.1381 & 0.0393 & 0.0658 & 0.0453 & 0.0550 & 0.0317 & 0.0515 & 0.0348 & 0.0422 \\
& Burn-down & 0.1032 & 0.1412 & 0.0998 & 0.1125 & 0.0369 & 0.0643 & 0.0443 & 0.0528 & 0.0310 & 0.0502 & 0.0331 & 0.0415 \\
\midrule

\multirow{2}{*}{\textbf{M3}}
& Gaussian & 0.0139 & 0.0162 & 0.0136 & 0.0144 & 0.0031 & 0.0049 & 0.0030 & 0.0041 & 0.0109 & 0.0164 & 0.0135 & 0.0187 \\
& Burn-down & 0.1352 & 0.1924 & 0.1447 & 0.1615 & 0.0414 & 0.0699 & 0.0466 & 0.0572 & 0.0382 & 0.0609 & 0.0392 & 0.0482 \\
\bottomrule 
\end{tabular}
}
\end{table*}

%% file: Tables/ablation_decay.tex
\begin{table*}
\centering
\caption{\R{The ablation results of different decay scheme}}
\label{tab:abla_decay} 

\resizebox{\linewidth}{!}{
\begin{tabular}{c cccccccccccc}
\toprule 
\multirow{2}{*}{Decay Scheme} & \multicolumn{4}{c}{\textbf{Gowalla}} & \multicolumn{4}{c}{\textbf{Yelp2018}} & \multicolumn{4}{c}{\textbf{Amazon-Book}} \\ 
\cmidrule[0.5pt](lr){2-5}\cmidrule[0.5pt](lr){6-9}\cmidrule[0.5pt](lr){10-13}
& R@10 & R@20 & N@10 & N@20 & R@10 & R@20 & N@10 & N@20 & R@10 & R@20 & N@10 & N@20  \\
\midrule 

\textbf{Exponential} & 0.1047 & 0.1506 & 0.1147 & 0.1275 & 0.0304 & 0.0521 & 0.0344 & 0.0423 & 0.0273 & 0.0464 & 0.0281 & 0.0358 \\
\textbf{Power} & 0.1317 & 0.1876 & 0.1400 & 0.1562 & 0.0370 & 0.0635 & 0.0372 & 0.0511 & 0.0359 & 0.0581 & 0.0368 & 0.0452 \\
\textbf{Linear} & 0.1303 & 0.1842 & 0.1407 & 0.1559 & 0.0392 & 0.0653 & 0.0431 & 0.0536 & 0.0324 & 0.0524 & 0.0330 & 0.0446 \\
\textbf{Gaussian} & 0.1315 & 0.1809 & 0.1427 & 0.1564 & 0.0397 & 0.0664 & 0.0437 & 0.0540 & 0.0357 & 0.0564 & 0.0373 & 0.0455 \\
\midrule
\textbf{Original} & 0.1352 & 0.1924 & 0.1447 & 0.1615 & 0.0414 & 0.0699 & 0.0466 & 0.0572 & 0.0382 & 0.0609 & 0.0392 & 0.0482 \\
\bottomrule 
\end{tabular}
}
\end{table*}

%% file: Tables/efficiency.tex
\begin{table}
\centering
\caption{\R{Training and sampling efficiency of \method{} and several representative baseline methods.}}
\label{tab:effiecency}
\begin{tabular}{c cccccc} 
\toprule
\multirow{2}{*}{Model} & \multicolumn{2}{c}{Amazon-Book} & \multicolumn{2}{c}{Yelp2018} & \multicolumn{2}{c}{Gowalla} \\
\cmidrule(lr){2-3} \cmidrule(lr){4-5} \cmidrule(lr){6-7}
 & Training
 & Inference & Training & Inference & Training & Inference \\
\midrule
LightGCN  & 14,820.7s & 131.6s & 6,731.4s & 82.3s & 5,453.5s & 70.1s \\
MacridVAE & 17,704.8s & 65.9s & 5,772.6s & 35.2s & 2,710.4s & 27.4s \\
DiffRec & 6,865.8s & 149.2s & 4,086.3s & 99.1s & 8,356.8s & 80.5s \\
DDRM      & 4,934.1s & 175.6s & 4,634.3s & 117.0s & 3,277.4s & 99.6s \\
FlowCF    & 3,235.6s & 92.8s & 1,791.3s & 44.9s & 2,112.6s & 35.2s \\
\midrule
\method{} & 3,786.8s & 135.9s & 3,314.0s & 81.8s & 2,346.9s & 74.6s \\
\bottomrule
\end{tabular}
\end{table}

%% file: Tables/indepth.tex
\begin{table*}
\centering
\caption{Model performance on different settings of optimization objectives and sampling approaches. By default, the training and sampling paradigm of \method{} can be regarded as instantaneous time steps for likelihood metric and binomial bridge method for the sampling process (the last line). 
}
\label{tab:indepth}
\setlength{\tabcolsep}{3pt}
\resizebox{\linewidth}{!}{
\begin{tabular}{cc cccccccccccc}
\toprule 
\multirow{2}{*}{Objective} & \multirow{2}{*}{Sampling} & \multicolumn{4}{c}{\textbf{Gowalla}} & \multicolumn{4}{c}{\textbf{Yelp2018}} & \multicolumn{4}{c}{\textbf{Amazon-Book}} \\ 
\cmidrule[0.5pt](lr){3-6}\cmidrule[0.5pt](lr){7-10}\cmidrule[0.5pt](lr){11-14}
& & R@10 & R@20 & N@10 & N@20 & R@10 & R@20 & N@10 & N@20 & R@10 & R@20 & N@10 & N@20  \\
\midrule 

Finite-time & Poisson & 0.0772 & 0.1404 & 0.0724 & 0.0992 & 0.0175 & 0.0336 & 0.0181 & 0.0247 & 0.0234 & 0.0392 & 0.0199 & 0.0281 \\
Instantaneous & Poisson & 0.0769 & 0.1368 & 0.0702 & 0.0949 & 0.0170 & 0.0331 & 0.0175 & 0.0249 & 0.0229 & 0.0387 & 0.0194 & 0.0275 \\
Finite-time & Binomial-bridge & 0.1347 & 0.1903 & 0.1444 & 0.1603 & 0.0410 & 0.0689 & 0.0463 & 0.0566 & 0.0372 & 0.0592 & 0.0390 & 0.0478 \\
Instantaneous & Binomial-bridge & 0.1351 &  0.1920 & 0.1450 & 0.1615 & 0.0413 & 0.0695 & 0.0463 & 0.0567 & 0.0374 & 0.0601 & 0.0387 & 0.0478 \\

\bottomrule 
\end{tabular}
}
\end{table*}

%% file: Content/6-conclusion.tex
\section{Conclusion}

In this work, we dive into the potential of diffusion processes for interacting systems and propose a custom-designed interests burn-down process to depict collaborative filtering tasks. The interests burn-down process models the personalized decay of user's interests towards candidate items. Leveraging this process and the reversed burn-up generative process, we propose an autoencoder-based CF model named \method{} that makes personalized recommendations based on interests diffusion process. Experimental results demonstrates the effectiveness and superiority of \method{} against other diffusion and AE-based methods. In the future, we aim to explore the applicability of the interests burn-down process in other recommendation scenarios to fully exploit its potential.